\documentclass[journal,twoside,web]{ieeecolor}
\usepackage{generic}
\usepackage{cite}
\usepackage{amsmath,amssymb,amsfonts}
\usepackage{algorithmic}
\usepackage{graphicx}
\usepackage{textcomp}
\usepackage{txfonts}
\usepackage{epsfig}
\usepackage{mathrsfs}
\usepackage{tipa}
\usepackage{booktabs}
\usepackage{caption}
\usepackage{setspace}
\usepackage{algorithm}
\usepackage{algorithmic}

\def\BibTeX{{\rm B\kern-.05em{\sc i\kern-.025em b}\kern-.08em
    T\kern-.1667em\lower.7ex\hbox{E}\kern-.125emX}}
\begin{document}
\title{Deep Reinforcement Learning Based Optimal Infinite-Horizon Control of Probabilistic Boolean Control Networks}
\author{Jingjie Ni, Fangfei Li, \IEEEmembership{Member, IEEE}, Zheng-Guang Wu, \IEEEmembership{Member, IEEE}
\thanks{This work was supported by the National Natural Science Foundation of China under Grants 62233005, 62173142, and the Programme of Introducing Talents of Discipline to Universities (the 111 Project) under Grant B17017. }
\thanks{Jingjie Ni is with the School of Mathematics, East China University of Science and Technology, Shanghai, 200237, P.R. China (email: nijingjie20000504@163.com).}
\thanks{Fangfei Li is with the School of Mathematics, East China University of Science and Technology, Shanghai, 200237, P.R. China. Meanwhile, she is also with the Key Laboratory of Smart Manufacturing in Energy Chemical Process, Ministry of Education, East China University of Science and Technology, Shanghai, 200237, P.R. China (email:
	li\_fangfei@163.com; lifangfei@ecust.edu.cn).}
\thanks{Zheng-Guang Wu is with the Institute of Cyber-Systems
\& Control, Zhejiang University, Hangzhou, Zhejiang, 310027, P.R. China (email:
nashwzhg@zju.edu.cn).}

}

\maketitle

\begin{abstract}
In this paper, a deep reinforcement learning based method is proposed to obtain optimal policies for optimal infinite-horizon control of probabilistic Boolean control networks (PBCNs). Compared with the existing literatures, the proposed method is model-free, namely, the system model and the initial states needn’t to be known. Meanwhile, it is suitable for large-scale PBCNs. First, we establish the connection between deep reinforcement learning and optimal infinite-horizon control, and structure the problem into the framework of the Markov decision process. Then, PBCNs are defined as large-scale or small-scale, depending on whether the memory of the action-values exceeds the RAM of the computer. Based on the newly introduced definition, Q-learning (QL) and double deep Q-network (DDQN) are applied to the optimal infinite-horizon control of small-scale and large-scale PBCNs, respectively. Meanwhile, the optimal state feedback controllers are designed. Finally, two examples are presented, which are a small-scale PBCN with 3 nodes, and  a large-scale one with 28 nodes. To verify the convergence of QL and DDQN, the optimal control policy and the optimal action-values, which are obtained from both the algorithms, are compared with the ones based on a model-based method named policy
iteration. Meanwhile, the performance of QL is compared with DDQN in the small-scale PBCN.
\end{abstract}


\section{Introduction}
\IEEEPARstart{T}{he} control of gene regulatory networks is a key problem in systems biology. In 1969, Kauffman\cite{BN1969} proposed Boolean networks (BNs) to model the dynamics of gene regulatory networks. In BNs, a Boolean variable ``0" or ``1" is used to represent whether a gene is transcribed or not. To describe the more complex behavior of gene regulatory networks, probabilistic Boolean networks (PBNs) have been proposed\cite{PBN2002}. PBNs are more general BNs that can describe random switching during gene regulation. Many gene regulatory networks have exogenous inputs, so PBNs are naturally extended to probabilistic Boolean control networks (PBCNs). Fundamental control problems of PBCNs have been carried out, such as controllability \cite{Controllability2011Li,Controllability2020Wu}, stabilization \cite{Stabilization2020HUANG,Stabilization2019Li,Stabilization2014Li}, finite-time stabilization \cite{FiStabilization2021Li,FiStabilization2021Liu,FiStabilization2019Guo}, observability \cite{Observability2020Fornasini,observabilityZHOU2019ZHOU}, output tracking \cite{output2019Yerudkar}, and finite-time output tracking\cite{Fioutput2021ZHANG}. Among them, optimal control of PBCNs is still an important problem, which is worth further study.

The objective of optimal control for a PBCN is to find optimal control policies, which minimize the cost-to-go function related to a control goal. Relevant research can be applied to therapeutic intervention\cite{QLoptimal,KLsamplebased}. For instance, the optimal policy obtained in \cite{QLoptimal} is successfully utilized in the melenoma gene-expression network to reduce the occurrence of undesired states. The other application of the study is engine control\cite{engine}, 

Optimal control of PBCNs can be divided into finite-horizon and infinite-horizon. The problems with finite-horizon have been considered in \cite{finite2010Liu,finite2015Wu,finite2018zhu}. Optimal infinite-horizon control of PBCNs is required for persistence problems. Unlike the one with finite-horizon, optimal infinite-horizon control of PBCNs does not necessarily have a limited cost-to-go function. The discounted factor has been introduced to guarantee the existence of solutions. Compared with optimal finite-horizon control, the design of the optimal infinite-horizon controller is more challenging due to increased computational complexity.

To solve optimal infinite-horizon control of PBCNs, policy iteration\cite{PI2006Pal,PI2018Wu,PI2019WU,PI2021Wu,STP2014Fornasini}, $Q$L\cite{QLoptimal}, and sampling-based path integral \cite{KLsamplebased} have been presented. Although the above methods are highly efficient, they have some limitations. Policy iteration is a matric-based method with the assumption that the system models of PBCNs are known. However, it cannot deal with large-scale PBCNs since the operation of gigantic metrics is infeasible. Meanwhile, the perfectly accurate systems models of PBCNs are impossible, and PBCNs with complex construction is difficult to be modeled due to computational complexity. To deal with modeling difficulty, model-free sampling-based path integral\cite{KLsamplebased} and $Q$L\cite{QLoptimal} are considered. Nevertheless, $Q$L\cite{QLoptimal} is not applicable to large-scale systems, and sampling-based path integral\cite{KLsamplebased} only provides an optimal policy according to a given initial state. 

To overcome the limitations mentioned above, DD$Q$N, a deep reinforcement method, is considered. The connection between reinforcement learning and optimal infinite-horizon control is revealed in \cite{bertsekas2019reinforcement}. A simple version of DD$Q$N is $Q$L \cite{QL}, which is a model-free algorithm in reinforcement learning with convergence guarantee \cite{QLcovergence}. Using $Q$L, an agent under the framework of the Markov decision process can learn optimal control policies through interaction with the environment \cite{Reinforcement,bertsekas2019reinforcement}. For PBCNs without known models, $Q$L presents a more direct way to obtain optimal policy since the modeling step is skipped. $Q$L has been widely used in control problems of gene regulatory networks, such as feedback stabilization of PBCNs \cite{QLStabilityPBCN2021,QLForest}, and optimal infinite-horizon control of PBCNs under state-flipped control\cite{QLoptimal}. For a gene regulatory network containing too many nodes, $Q$L is no longer valid due to $curse$ $of$ $dimensionality$, and deep $Q$-network \cite{2015Human} in deep reinforcement learning needs to be considered. Deep $Q$-network is the combination of $Q$L and deep learning. The advantage of deep $Q$-network over $Q$L is its ability to solve problems with large state space. DD$Q$N is an advanced version of deep $Q$-network\cite{2016DDQN}, which decreases the overestimation in action-values of deep $Q$-network. DD$Q$N only needs to add a network on the basis of deep $Q$-network, which is relatively simple compared with other methods to improve the performance of deep $Q$-network. Therefore, DD$Q$N has been applied in various studies, such as controllability of PBCNs under external perturbations\cite{DDQNcontrollability}, and output tracking of PBCNs \cite{DDQNoutputTracking2020}. Although many control problems of PBCNs based on $Q$L and DD$Q$N have been investigated, most of them focus on controllability and stabilization \cite{QLStabilityPBCN2021,QLForest,DDQNcontrollability,DDQNoutputTracking2020}. It is worth noting that the optimal infinite-horizon control of PBCN based on deep reinforcement learning is still an open problem.

Taken together, in this paper we utilize model-free $Q$L and DD$Q$N to obtain optimal control policies for small-scale and large-scale PBCNs. The main contributions of our paper are listed as follows.
\begin{enumerate}
	\item  In Theorem 1, we prove that reinforcement learning provides an optimization framework to solve the infinite horizon problem, which can be easily framed as a minimization problem. 
	\item $Q$L and DD$Q$N for optimal infinite-horizon control of PBCNs are proposed. Unlike \cite{PI2006Pal,PI2018Wu,PI2019WU,PI2021Wu,STP2014Fornasini} and \cite{KLsamplebased}, our method does not need the knowledge of system models or initial states.
	\item Optimal infinite-horizon control of large-scale PBCNs is solved in this paper, which cannot be solved by using model-based methods proposed in \cite{PI2006Pal,PI2018Wu,PI2019WU,PI2021Wu,STP2014Fornasini}, as well as $Q$L in \cite{QLoptimal}.
\end{enumerate}

The framework of this paper is organized as follows. In
Section \uppercase\expandafter{\romannumeral2}, we introduce the basic concepts of the Markov decision process, $Q$L, and DD$Q$N. In
Section \uppercase\expandafter{\romannumeral3}, we give the system models of PBCN. Then, we structure optimal infinite-horizon control of PBCNs into the framework of the Markov decision process, and reveal the relationship between deep reinforcement learning and optimal control. Next, we propose optimal infinite-horizon control of small-scale and large-scale PBCNs using $Q$L and DD$Q$N, respectively. In
Section \uppercase\expandafter{\romannumeral4}, a small-scale PBCN with 3 nodes and a large-scale one with 28 nodes are considered to verify the proposed method. Finally, the conclusions are given in Section \uppercase\expandafter{\romannumeral5}.

$\mathbf{Notations:}\ \mathbb{Z^+}$, $\mathbb{R}^{+} $, $\mathbb{R}$ denotes the sets of nonnegative integers, nonnegative real numbers, and real numbers, respectively. $\mathbb{E}[\cdot] $ is the expected value operator. var$[\cdot] $ is the variance. $ \Pr\left\{A\mid B\right\} $ is the probability of the event $A$ under the condition of the event $B$. For a set $D$, $|D|$ is the number of elements. There are three basic operations on Boolean variables, which are ``not", ``and" and ``or", expressed as $\neg$, $\wedge$, and $\vee$, respectively. $\mathcal{B}:=\{0,1\}$ is the Boolean domain, and $\mathcal{B}^{n}:=\underbrace{\mathcal{B} \times \ldots \times \mathcal{B}}_{n}$. 

\section{Preliminaries}
In this section, we introduce deep reinforcement learning, which lays the foundation for solving optimal infinite-horizon control of PBCNs. First, we introduce the framework of reinforcement learning, namely the Markov decision process. Then, we introduce two algorithms named $Q$L and DD$Q$N.

\subsection{Markov Decision Process}
\begin{figure}[!t]
	\centerline{\includegraphics[width=\columnwidth]{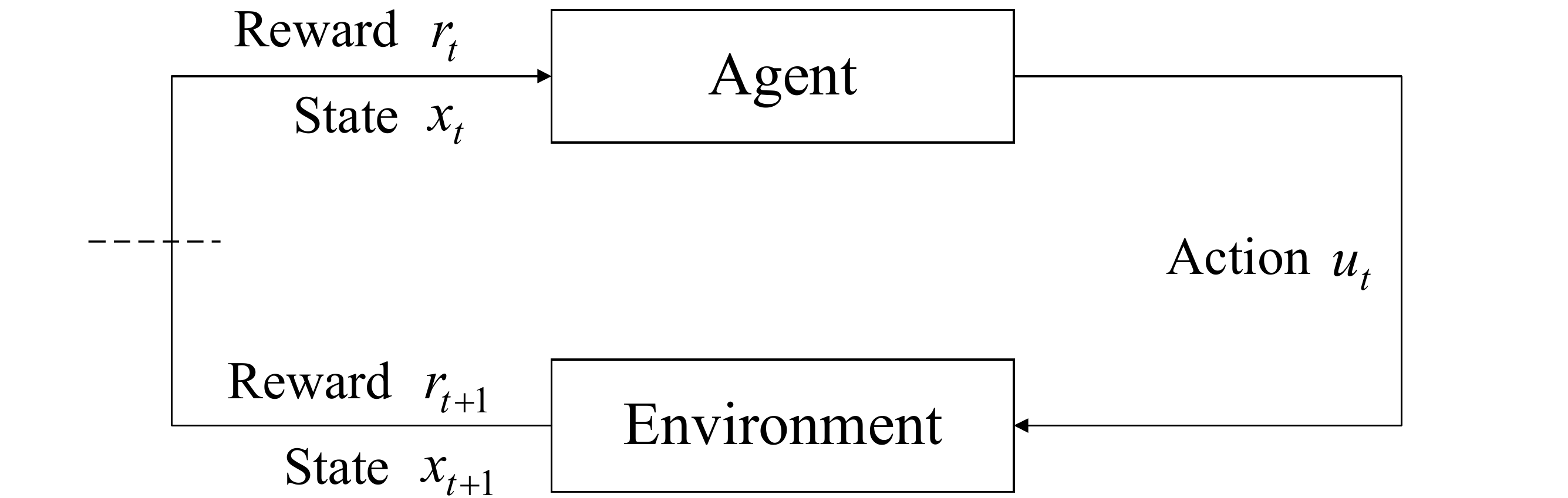}}
	\caption{Markov decision process}
	\label{figMDP}
\end{figure}

The schematic diagram of the Markov decision process is shown in Figure \ref{figMDP}. Let a quintuple $(\mathbf{X}, \mathbf{U}, \gamma, \mathbf{P}, \mathbf{R})$ represent the Markov decision process, where $\mathbf{X}=\{x_t,t\in\mathbb{Z^+}\} $ is the state space, $\mathbf{U}=\{u_t,t\in\mathbb{Z^+}\} $ is the action space, $0\leq\gamma <1 $ is the discount factor, $\mathbf{P}_{x_t}^{x_{t+1}}(u_t)=\Pr\left\{x_{t+1} \mid x_t, u_t\right\}$ is the state-transition probability from state $x_t$ to $x_{t+1}$ when action $u_t$ is taken, and $\mathbf{R}_{x_t}^{x_{t+1}}(u_t)=\mathbb{E}\left[r_{t+1} \mid x_t, u_t\right]$ with $r_{t+1}=r_{t+1}(x_t,u_t)$ is the expected reward. The Markov decision process is the framework for an agent to find an optimal policy through interaction with the environment. At each time step $ t\in \mathbb{Z^+}$, the agent observes a state $x_{t}$ and selects an action $u_t$, according to the policy $\pi:x_t\rightarrow u_t, \forall t\in\mathbb{Z^+}$. Then, the environment gives a reward $r_{t+1}$ and a new state $x_{t+1}$. The agent obtains $r_{t+1}$ which reveals the advantage to take $u_t$ at $x_{t}$, and then updates $\pi$. 

Define $G_t=\sum_{i=t+1}^{\infty} \gamma^{i-t-1} r_t$ as the return. The goal of the agent is to find the optimal deterministic policy $\pi^*:x_t\rightarrow u_t, \forall t\in\mathbb{Z^+}$, under which the expected return $\mathbb{E}_\pi[G_t], \forall t\in\mathbb{Z^+}$ is maximized. $q_\pi(x_t,u_t)$ is the value of taking the action $u_t$ at the state $x_t$ and thereby following the policy $\pi$, which is a detailed version of $\mathbb{E}_\pi[G_t]$:
\begin{equation}
	q_\pi(x_t,u_t) =\mathbb{E}_\pi[G_t|x_t,u_t],
	\label{eqactionvalue}
\end{equation}
and satisfies the Bellman equation:
\begin{equation}
	q_\pi(x_t,u_t)=\sum\limits_{x_{t+1}\in\mathbf{X}}\mathbf{P}_{x_t}^{x_{t+1}}(u_t)[\mathbf{R}_{x_t}^{x_{t+1}}(u_t)+\gamma q_\pi(x_{t+1},\pi(x_{t+1}))].
	\label{eqBL}
\end{equation}
Equation (\ref{eqBL}) reveals the recursion of $q_\pi(x_t,u_t)$, which is critical for the algorithms introduced in the following subsection. 

The optimal action-value is defined as follows:
\begin{equation}
	q^*(x_t,u_t) = \max\limits_{\pi\in\Pi} q_\pi(x_t,u_t),\forall x_t\in \mathbf{X},\forall u_t\in \mathbf{U},
\end{equation}
where $\Pi$ is the set of all admissible policies. Based on $q^*(x_t,u_t)$, $\pi^*$ can be obtained as follows:
\begin{equation}
	\pi^*(x_t) = \arg\max\limits_{u_t\in\mathbf{U}} q^*(x_t,u_t),\forall x_t\in \mathbf{X}.
	\label{eqoptpolicy}
\end{equation}

For problems under the framework of the Markov decision process, $\pi^* $ can be obtained through many algorithms, e.g., policy iteration. However, the premise
of these algorithms is stringent to optimal infinite-horizon control of PBCNs. In particular, the model including $\mathbf{P}$ should be known, whereas modeling PBCNs is difficult. Therefore, we introduce a model-free technique, namely $Q$L.

\subsection{Q-Learning}
$Q$L is a classical algorithm in reinforcement learning. For a problem in the framework of the Markov decision process with an unknown model, an agent can obtain $\pi^*$ through interaction with the environment using $Q$L.

\begin{figure}[!t]
	\centerline{\includegraphics[width=\columnwidth]{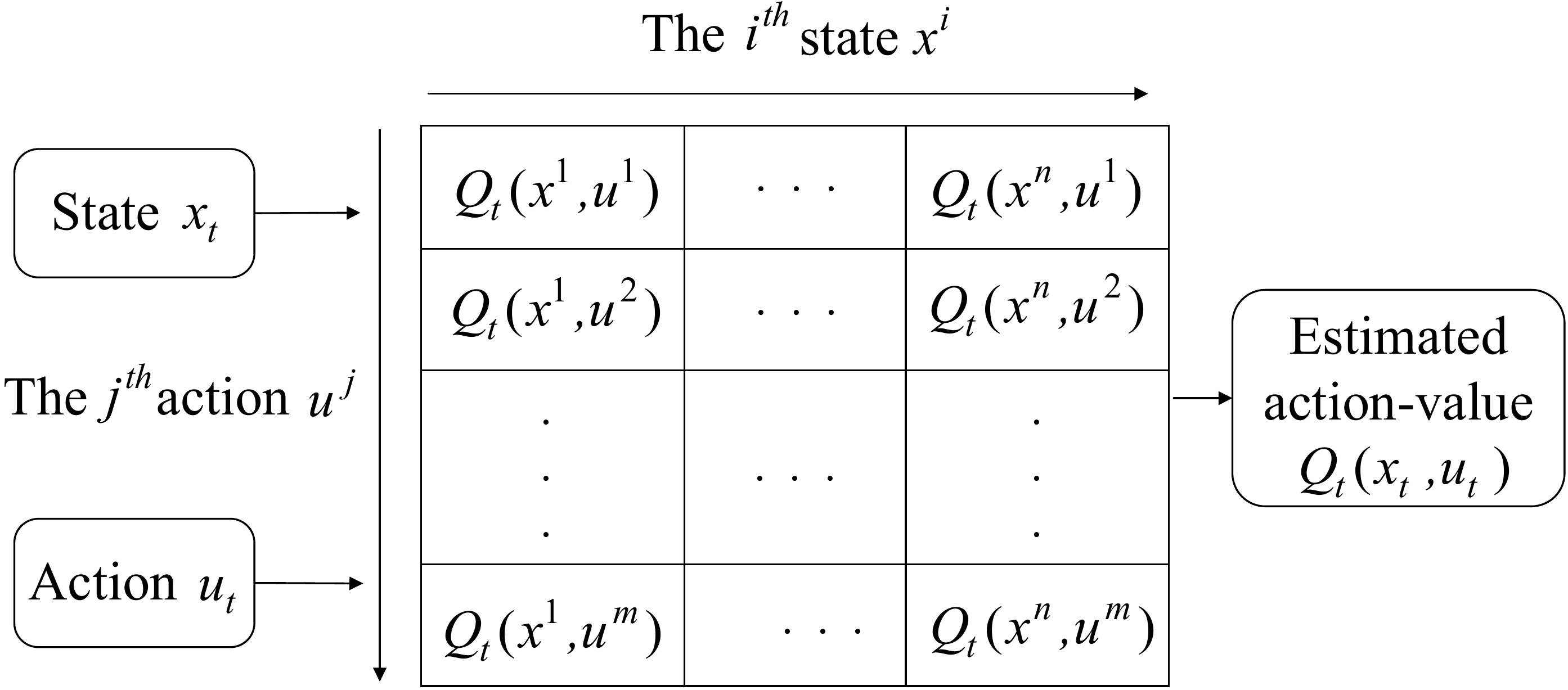}}
	\caption{$Q$-table}
	\label{figQtable}
\end{figure}

As shown in Figure \ref{figQtable}, the $Q$-table is used to record each $Q_{t}(x_t, u_t)$, which is the estimate of $q^*(x_t, u_t)$. At each time step $t$, the $Q$-table updates once, using the rule given as follows:
\begin{equation}
	Q_{t+1}(x, u)=\left\{
	\begin{aligned}
		&(1-\alpha_t)Q_{t}(x, u)+\alpha_t TDE_{t},&&\text{if}(x, u)=\left(x_{t}, u_{t}\right),\\
		&Q_{t}(x, u), &&\text{else},
	\end{aligned}\right.
	\label{eqQupdate}
\end{equation}
where $\alpha_t \in (0,1], t\in \mathbb{Z^+}$ is the learning rate, and $TDE_{t}$ is the temporal-difference (TD) error:
\begin{equation}
	TDE_{t}= r_{t+1}+\gamma\max\limits_{u_{t+1}\in\mathbf{U}}Q_t(x_{t+1},u_{t+1})-Q_t(x_t,u_t) .
\end{equation}
$\epsilon $-greedy method is used for an agent to select an action:
\begin{equation}
	u_{t}=\left\{\begin{aligned}
		&\arg \max\limits_{u_t \in \mathbf{U}} Q_{t}\left(x_{t}, u_t\right), &&P=1-\epsilon,\\
		&\operatorname{rand}(\mathbf{U}) ,&&P=\epsilon,\\
	\end{aligned}\right. 
\end{equation}
where $\operatorname{rand}(\mathbf{U}) $ represents an action $u_t$ which is randomly selected from the action space $\mathbf{U}$. Following the $\epsilon $-greedy method, an agent not only tests new actions to find a better policy but also develops the current optimal policy.

Under certain conditions, $Q_t(x_{t},u_{t})$ converges to the fixed point $q^*(x_t,u_t)$ with probability one \cite{QLcovergence}:
\begin{equation}
	\label{eqmaxQ}
	\lim_{t\rightarrow\infty}Q_t(x_t,u_t) = q^*(x_t,u_t),\forall x_t\in \mathbf{X},\forall u_t\in \mathbf{U}.
\end{equation}
Substitute (\ref{eqmaxQ}) into (\ref{eqoptpolicy}), the optimal policy $\pi^*$ is obtained as follows:
\begin{equation}
	\pi^{*}\left(x_{t}\right)=\arg \underset{u_t \in \mathbf{U}}{\max }\lim_{t\rightarrow\infty} Q_t\left(x_{t}, u_t\right) ,\forall x_{t} \in \mathbf{X}.
\end{equation}

The limitation of $Q$L is that it cannot solve problems with large state space. In particular, the operation of $Q$L is based on a $Q$-table, which requires enough memory to store $|\mathbf{X}|\times|\mathbf{U}| $ values. When the state space $\mathbf{X}$ is so large that the memory of $|\mathbf{X}|\times|\mathbf{U}| $ values exceeds the capabilities of modern computers, $Q$L is no longer applicable. For these problems, we introduce DD$Q$N.

\subsection{Double Deep Q-Network}
DD$Q$N is an algorithm in deep reinforcement learning, which is the combination of deep learning and $Q$L. The difference between DD$Q$N and $Q$L lies in the expression of estimated action-values. $Q$L uses a $Q$-table to represent estimated action-values, while DD$Q$N uses function approximation. Due to the difference, DD$Q$N no longer needs a large memory to store $|\mathbf{X}|\times|\mathbf{U}| $ values. Thus, it is suitable for problems with large state space.

\begin{figure}[!t]
	\centerline{\includegraphics[width=\columnwidth]{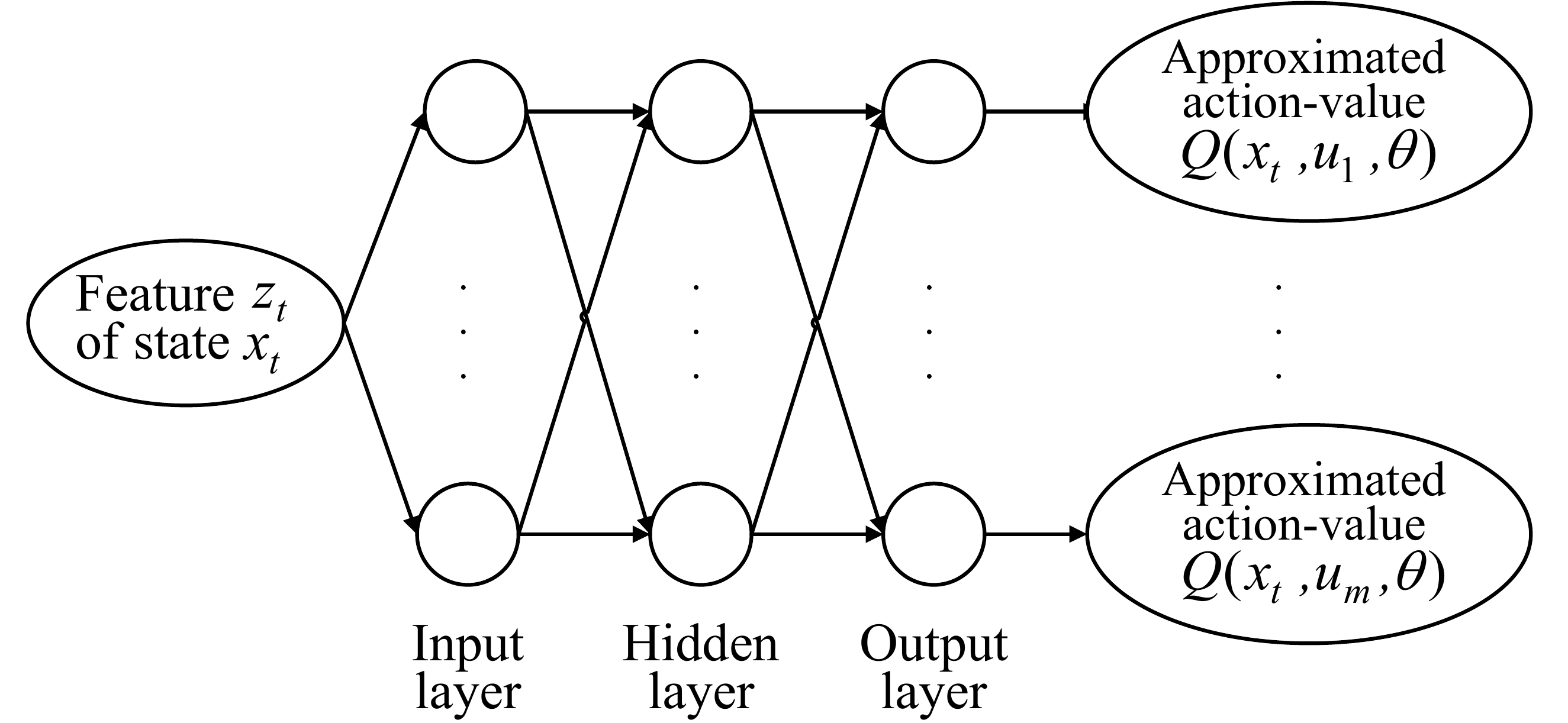}}
	\caption{Artificial neural network with one hidden layer}
	\label{figAnn}
\end{figure}
Artificial neural networks are used for function approximation. A good property of artificial neural networks is that they can fit any function. The structure of the artificial neural network with a hidden layer is shown in Figure \ref{figAnn}. Given the feature $z_t$ of the state $x_t$, the outputs of the artificial neural network are the approximated action-values of taking $u_i,i=1,\dots,m,$ at $x_t$:
\begin{multline}
	Q(x_t,u_i,\theta) =F_l(w_l(\dots F_2(w_2F_1(w_1z_t+b_1)\\
	+b_2)\dots)+b_l),i=1,\dots,m,
\end{multline}
where $l$ represents the number of layers. $F_j$, $w_j\in\mathbb{R}^{n_{j}}\times\mathbb{R}^{n_{j+1}} $, and $b_i\in\mathbb{R}^{n_{j}}$ represent the activation function, the weight matrix, and the bias vector of the $j^{th}$ layer respectively. Besides, $\theta=\{w_j,b_j\}_{j=1}^{l}$ is the set of parameters.

The objective of the artificial neural network is to fit the optimal action-value function, and then obtain the optimal policy:
\begin{equation}
	\begin{aligned}
		Q(x_t,u_t,\theta)\approx q^*(x_t,u_t),\ \forall x_t\in \mathbf{X},\\
		\pi^*(x_t)=\arg \max\limits_{u_t\in\mathbf{U}}Q(x_t,u_t,\theta),\ \forall x_t\in \mathbf{X}.
	\end{aligned}
\end{equation}

At each time step $t$, the artificial neural network updates $\theta $ to reduce the loss function $L(\theta)$, which is given as follows:
\begin{equation}
	L(\theta)=\frac{1}{\mathcal{M}}\sum\limits_i^\mathcal{M}(y_i-Q(x_i,u_i,\theta))^2.
	\label{eqloss}
\end{equation}
In (\ref{eqloss}), mini-batch size $\mathcal{M}$ is the number of samples, and $y_i $ represents the $i^{th}$ target value:
\begin{equation}
	\begin{aligned}
		&y_i=r_i+\gamma Q(x_i',\arg\max\limits_{u\in\mathbf{U}}Q(x_i',u,\theta),\varphi),
	\end{aligned}
	\label{eqyi}
\end{equation}
where $x_i'$ is the subsequent state of $x_i$, and $\varphi$ is the target network. Polyak averaging is used for updating $\varphi$, namely $\varphi=\tau\varphi+(1-\tau)\theta$, where $0\leq\tau\leq 1$ is the learning rate of $\varphi$.

The selection of $y_i, i=1,\dots, \mathcal{M}$ is based on experience replay, which is a process of random sampling from the replay memory. The replay memory $D_k = \{e_1,\dots,e_k\} $ stores $k$ latest Markov decision process sequences, where $k$ represents the capacity of $D_k$. The purpose of experience replay is to achieve stability by breaking the temporal dependency among samples.

The artificial neural network reduces the loss function $L(\theta) $ (\ref{eqloss}) by updating the parameters $\theta $ with gradient descent:
\begin{equation}
	\theta=\theta-\beta \nabla_\theta L(\theta),
\end{equation}
where $0<\beta\leq1 $ is the learning rate, and $ \nabla_\theta L(\theta)$ is the gradient of  $L(\theta) $ to $\theta$.

\section{Optimal Control of PBCNs Using Deep Reinforcement Learning}
In this section, we discuss optimal infinite-horizon control of PBCNs using deep reinforcement learning. The advantage of deep reinforcement learning in the problem lies in its model-free characteristic, which solves the difficulty of modeling PBCNs. First, we present the system model of PBCNs. Then, we use a model-free approach based on deep reinforcement learning to solve the problem. Specifically, we establish the connection between action-value functions in deep reinforcement learning and cost-to-go
functions in traditional optimal infinite-horizon control, and structure the problem into the framework of the Markov decision process. Then, we propose optimal infinite-horizon control of PBCNs using $Q$L and DD$Q$N. The details of the proposed algorithms, such as the setting of parameters and the structure of artificial neural networks, are explained. Finally, we discuss the computational complexity of $Q$L and DD$Q$N, and then present the applicability of these two algorithms.

\subsection{System Models for PBCNs}

A PBCN with $n$ nodes and $m$ control inputs is defined as follows:
\begin{equation}
	\left\{\begin{aligned}
		&x_{1}(t+1)=f_{1}^{(j)}\left(x_{1}(t), \ldots, x_{n}(t), u_{1}(t), \ldots, u_{m}(t)\right), \\
		&x_{2}(t+1)=f_{2}^{(j)}\left(x_{1}(t), \ldots, x_{n}(t), u_{1}(t), \ldots, u_{m}(t)\right), \\
		&\qquad\qquad\vdots \\
		&x_{n}(t+1)=f_{n}^{(j)}\left(x_{1}(t), \ldots, x_{n}(t), u_{1}(t), \ldots, u_{m}(t)\right),
	\end{aligned}\right.  \ t\in \mathbb{Z^{+}},
	\label{eqPBCN}
\end{equation}
where $x_i(t)\in \mathcal{B}, i \in \{1,...,n\}$ represents the $i^{th}$ node at the time step $t$, and $u_j(t)\in \mathcal{B}, j\in \{1,...,m\}$ represents the $j^{th}$ control input at the time step $t$. All nodes at the time step $t$ are represented by $x(t)=\left(x_{1}(t), \ldots, x_{n}(t) \right) \in \mathcal{B}^{n} $. Similarly, all control inputs at the time step $t$ are represented by $u(t)=\left(u_{1}(t), \ldots, u_{m}(t) \right) \in \mathcal{B}^{m} $. Taking $x(t)\times u(t)$ as independent variables, and $x_i(t+1) $ as dependent variable, the logical function is described as $f_{i} ^{(j)}\in \mathcal{F}_{i}=\left\{f_{i}^{1}, f_{i}^{2}, \ldots, f_{i}^{l_ i}\right\}: \mathcal{B}^{n+m} \rightarrow \mathcal{B} , j \in \{1,...,l_i\}$. A logical function $f_{i} ^{(j)}$ is randomly selected from the set of logical functions $\left\{f_{i}^{1}, f_{i}^{2}, \ldots, f_{i}^{l_ i}\right\}$ with the probability $\left\{\mathrm{P}_{i}^{1}, \mathrm{P}_{i}^{2}, \ldots, \mathrm{P}_{i}^{l_{i}}\right\} $, where $\sum_{j=1}^{l_{i}} \mathrm{P}_{i}^{j}=1 $ and $\mathrm{P}_{i}^{j} \geq 0 $. 

$\mathbf{Remark\ 1:}$ PBCNs, which can describe the random
switching behaviors, are more general versions of Boolean
control networks. When $l_i=1,\ i=1,...,n$, PBCNs degenerate into Boolean control networks.

\subsection{Optimal Infinite-horizon Control of PBCNs in Markov decision process}
For optimal infinite-horizon control of PBCNs, we aim to find the optimal deterministic control policy $\tilde{\pi}^*:x(t)\rightarrow u(t), \forall x(t) \in \mathcal{B}^n$ which minimizes the cost-to-go function:
\begin{equation}
	J_{\tilde{\pi}}(x(t))=\lim_{N\rightarrow \infty}\mathbb{E}_{\tilde{\pi}}\left[\sum_{i=t}^{N} \gamma ^{i-t}l_{i+1}|x(t)\right],\forall x(t) \in \mathcal{B}^n,
	\label{eqcosttogo}
\end{equation}
where $l_{i+1}=l_{i+1}\left(x(i),u(i)\right)\in \mathbb{R}^+$ is the cost. The discount factor $\gamma$ ensures the finiteness of $J_{\tilde{\pi}}(x(t))$ for infinite-horizon, which prevents ill-posed optimization problems.

$\mathbf{Lemma\ 1:}$ Consider (\ref{eqcosttogo}), if there exists a constant $M\in \mathbb{R}^+$ such that $0\leq l_{i+1}\leq M,\forall i \in \mathbb{Z^{+}}$, then $J_{\tilde{\pi}}(x(t))$ is bounded.

$\mathbf{Proof:}$ Notice that there exists a constant $M \in \mathbb{R}^{+}$ such that $0\leq l_{i+1}\leq M, \forall i \in \mathbb{Z^{+}} $, consequently, we have:
\begin{equation}
	\begin{aligned}
		J_{\tilde{\pi}}(x(t))&=\lim_{N\rightarrow \infty}\mathbb{E}_{\tilde{\pi}}\left[\sum_{i=t}^{N} \gamma ^{i-t}l_{i+1}|x(t)\right] \\
		&=\mathbb{E}_{\tilde{\pi}}\left[\sum_{i=t}^{\infty} \gamma ^{i-t}l_{i+1}|x(t)\right] \\
		&=\sum_{i=t}^{\infty} \gamma ^{i-t}\mathbb{E}_{\tilde{\pi}}\left[l_{i+1}|x(t)\right]\\
		&\leq \sum_{i=t}^{\infty} \gamma ^{i-t}M\\
		&=\frac{M}{1-\gamma},
		\nonumber
	\end{aligned}
\end{equation} 
where the second equality and the third one hold according to Theorem 1.4.44 and Corollary  1.4.46 of \cite{tao2011introduction}, respectively. Thus, $J_{\tilde{\pi}}(x(t))$ is bounded. $\hfill\blacksquare$

The introduction of $\gamma $ brings new meaning to $J_{\tilde{\pi}}(x(t))$. $J_{\tilde{\pi}}(x(t))$ gives higher weights to recent costs, and lower weights to future costs. This definition is consistent with most control goals of PBCNs. For instance, optimal infinite-horizon control of PBCNs can be applied to cancer therapy, where the initial condition of patients is more concerned with the consideration of life expectancy.

\begin{figure}[!t]
	\centerline{\includegraphics[width=\columnwidth]{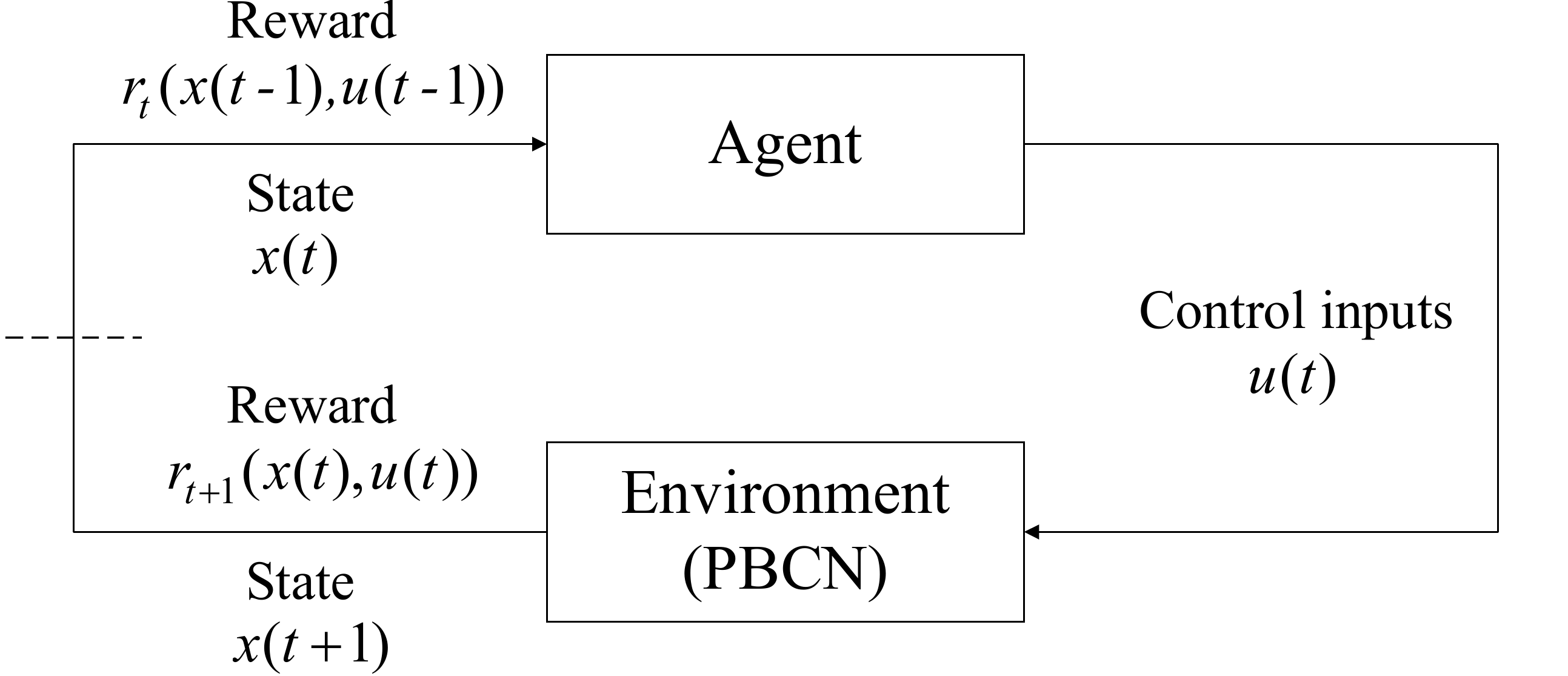}}
	\caption{Markov decision process for PBCNs}
	\label{figMDPPBCN}
\end{figure}

The premise for using deep reinforcement learning in optimal infinite-horizon control of PBCNs is structuring the problem into the framework of the Markov decision process. We represent the Markov decision process for optimal infinite-horizon control of PBCNs by the quintuple $(\mathcal{B}^n, \mathcal{B}^m, \gamma,\mathbf{P}, \mathbf{R}) $. The details of the Markov decision process are shown in Figure \ref{figMDPPBCN}. The environment is defined as the PBCN, and the agent is the machine that learns and provides the policy. The state space is represented by $\mathcal{B}^n $. A state is defined as $x_t=x(t) $. The action space is represented by $\mathcal{B}^m$. An action is defined as $u_t=u(t) $. The discount factor $ \gamma $ is valued in $[0,1)$, according to the importance of the future reward. The state-transition probability $ \mathbf{P} $ is derived from the dynamics of PBCNs. The expected reward $ \mathbf{R} $ is the expectation of $ r_t $.

It is assumed that the agent knows the state dimension $n$ and the action dimension $m$, but does not know $\mathbf{P}$ and $\mathbf{R}$. In other words, the agent only knows the number of nodes $n$ and the number of control inputs $m$, but does not know the dynamics of the PBCN (\ref{eqPBCN}). Meanwhile, at each time step $t$, the agent is assumed to receive $x_t$ and $r_{t}$ given by the PBCN, and the PBCN is also assumed to receive $u_t$ given by the agent. Each interaction between the agent and the environment helps the agent further understand $\mathbf{P}$ and $\mathbf{R}$, which is reflected in a better estimation of $q^*(x_t,u_t)$ and a more in-depth knowledge of $\pi^*$.

Notice that the expression of the optimal policy in the Markov decision process is different from that in traditional optimal control. To be specific, $\pi^*$ is defined as the one maximizes $q_\pi(x_t,u_t)$ in Markov decision process, whereas $\tilde{\pi}^*$ is defined as the one minimizes $J_{\tilde{\pi}}(x_t)$ in traditional optimal control. Next, we will show that $\pi^*$ is equivalent to $\tilde{\pi}^*$ under certain reward settings.

$\mathbf{Theorem\ 1:}$ Set $r_{i+1}=C_1l_{i+1}+C_2, \forall i \in \mathbb{Z}^{+}$, where $C_1<0$ and $C_2$ are two constants, then ${\pi}^*$ is equivalent to $\tilde{\pi}^*$.

$\mathbf{Proof:}$ First, we show that $\pi^*$ minimizes $J_{\pi}(x_t), \forall x_t \in \mathbf{X}$. Since there exist two constants $C_1<0$ and $C_2$ such that $r_{i+1}=C_1l_{i+1}+C_2, \forall i \in \mathbb{Z}^{+}$, it is easy to obtain $l_{i+1}=\frac{r_{i+1}-C_2}{C_1}, \forall i \in \mathbb{Z}^{+}$. Substitute $l_{i+1}=\frac{r_{i+1}-C_2}{C_1}$ into $J_{\pi}(x_t)=\lim_{N\rightarrow \infty}\mathbb{E}_\pi\left[\sum_{i=t}^{N} \gamma ^{i-t}l_{i+1}|x_t\right]$, then we obtain 
\begin{equation}
	\begin{aligned}
		J_{\pi}(x_t)&=\mathbb{E}_\pi\left[\sum_{i=t}^{\infty} \gamma ^{i-t}\frac{r_{i+1}-C_2}{C_1}|x_t\right]
		\\
		&=\frac{1}{C_1}\mathbb{E}_\pi\left[\sum_{i=t}^{\infty} \gamma ^{i-t}r_{i+1}|x_t\right]-\frac{C_2}{C_1}\sum_{i=t}^{\infty}\gamma^{i-t}
		\\
		&=\frac{1}{C_1}\mathbb{E}_\pi\left[\sum_{i=t}^{\infty} \gamma ^{i-t}{r_{i+1}}|x_t,\pi(x_t)\right]+\frac{C_2}{(1-\gamma)C_1}
		\\
		&=\frac{1}{C_1}q_\pi(x_t,\pi(x_t))+\frac{C_2}{(1-\gamma)C_1}.
		\nonumber
	\end{aligned}
\end{equation} 
Recall that $\pi^*$ maximizes $q_\pi(x_t,u_t), \forall x_t \in \mathbf{X}, \forall u_t \in \mathbf{U}$. Meanwhile, equation (\ref{eqoptpolicy}) indicates that $q_{\pi^*}(x_t,\pi^*(x_t)) = \max\limits_{\forall u_t} q^*(x_t,u_t)$. Thus, $\pi^*$ maximizes $q_\pi(x_t,\pi(x_t)), \forall x_t \in \mathbf{X}$. Notice that only $\frac{1}{C_1}q_\pi(x_t,\pi(x_t))$ depends on $\pi$ in the last equality, where $C_1<0$. Hence, we can conclude that $\pi^*$ minimizes $J_{\pi}(x_t), \forall x_t \in \mathbf{X}$. 

Next, we show that $\tilde{\pi}^*$ maximizes $q_{\pi}(x_t,u_t), \forall x_t \in \mathbf{X}, \forall u_t \in \mathbf{U}$. 
Substitute $r_{i+1}=C_1l_{i+1}+C_2$ into $q_{\pi}(x_t,u_t)=\mathbb{E}_\pi\left[\sum_{i=t}^{\infty} \gamma ^{i-t}{r_{i+1}}|x_t,u_t\right]$, then we obtain 
\begin{equation}
	\begin{aligned}
		q_{\pi}(x_t,u_t)&=\mathbb{E}_\pi\left[\sum_{i=t}^{\infty} \gamma ^{i-t}(C_1l_{i+1}+C_2)|x_t,u_t\right]
		\\
		&=C_1\mathbb{E}_\pi\left[\sum_{i=t}^{\infty} \gamma ^{i-t}l_{i+1}|x_t,u_t\right]+C_2\sum_{i=t}^{\infty}\gamma^{i-t}
		\\
		&=C_1\sum_{\forall x_{t+1}\in \mathbf{X}}\mathbf{P}_{x_t}^{x_{t+1}}(u_t)\mathbb{E}_\pi\left[\sum_{i=t+1}^{\infty} \gamma ^{i-t}{l_{i+1}}|x_{t+1}\right]\\
		&\ \ +C_1\mathbb{E}\left[l_{t+1}|x_t,u_t\right]+\frac{C_2}{1-\gamma}
		\\
		&=C_1\sum_{\forall x_{t+1}\in \mathbf{X}}\mathbf{P}_{x_t}^{x_{t+1}}(u_t)\lim_{N\rightarrow \infty}\mathbb{E}_\pi\left[\sum_{i=t+1}^{N} \gamma ^{i-t}l_{i+1}|x_{t+1}\right]\\
		&\ \ +C_1\mathbb{E}\left[l_{t+1}|x_t,u_t\right]+\frac{C_2}{1-\gamma}
		\\
		&=C_1\sum_{\forall x_{t+1}\in \mathbf{X}}\mathbf{P}_{x_t}^{x_{t+1}}(u_t)J_\pi(x_{t+1})+C_1\mathbb{E}\left[l_{t+1}|x_t,u_t\right]+\frac{C_2}{1-\gamma}.
		\nonumber
	\end{aligned}
\end{equation} 
Recall that $\tilde{\pi}^*$ minimizes $J_\pi(x_{t+1}), \forall x_{t+1} \in \mathbf{X}$. Notice that only $C_1\sum_{\forall x_{t+1}\in \mathbf{X}}\mathbf{P}_{x_t}^{x_{t+1}}(u_t)J_\pi(x_{t+1})$ depends on $\pi$ in the last equality, where $C_1<0$. Hence, $\tilde{\pi}^*$ maximizes $q_{\pi}(x_t,u_t), \forall x_t \in \mathbf{X}, \forall u_t \in \mathbf{U}$. Now, it can be concluded that ${\pi}^*$ is equivalent to $\tilde{\pi}^*$.
$\hfill\blacksquare$

Theorem 1 shows how to design rewards such that the traditional optimal control problem can be solved in the framework of the Markov decision process, where $\pi^*$ is equivalent to $\tilde{\pi}^*$. In the following, the algorithms for obtaining $\pi^*$, namely, $Q$L and DD$Q$N, are given.

\subsection{Optimal Control of Small-scale PBCNs Using QL}
For optimal infinite-horizon control of a small-scale PBCN, $Q$L is used to obtain an optimal policy $\pi^* $. We define PBCNs as small-scale if and only if the memory of the action-values is within the RAM of the computer. Since a value takes $2^3$ bytes, and 1 byte equals $2^{-10}$ GB, a PBCN is regarded as small-scale when $2^{m+n-7}$ is larger than the RAM (GB).
For these problems, $Q$L has two advantages. Firstly, $Q$L is a model-free algorithm, which resolves the difficulty of modeling PBCNs. Secondly, $Q$L has high computational efficiency, which ensures that an optimal control policy $\pi^* $ can be obtained in a short time.

\begin{figure}[!t]
	\centerline{\includegraphics[width=\columnwidth]{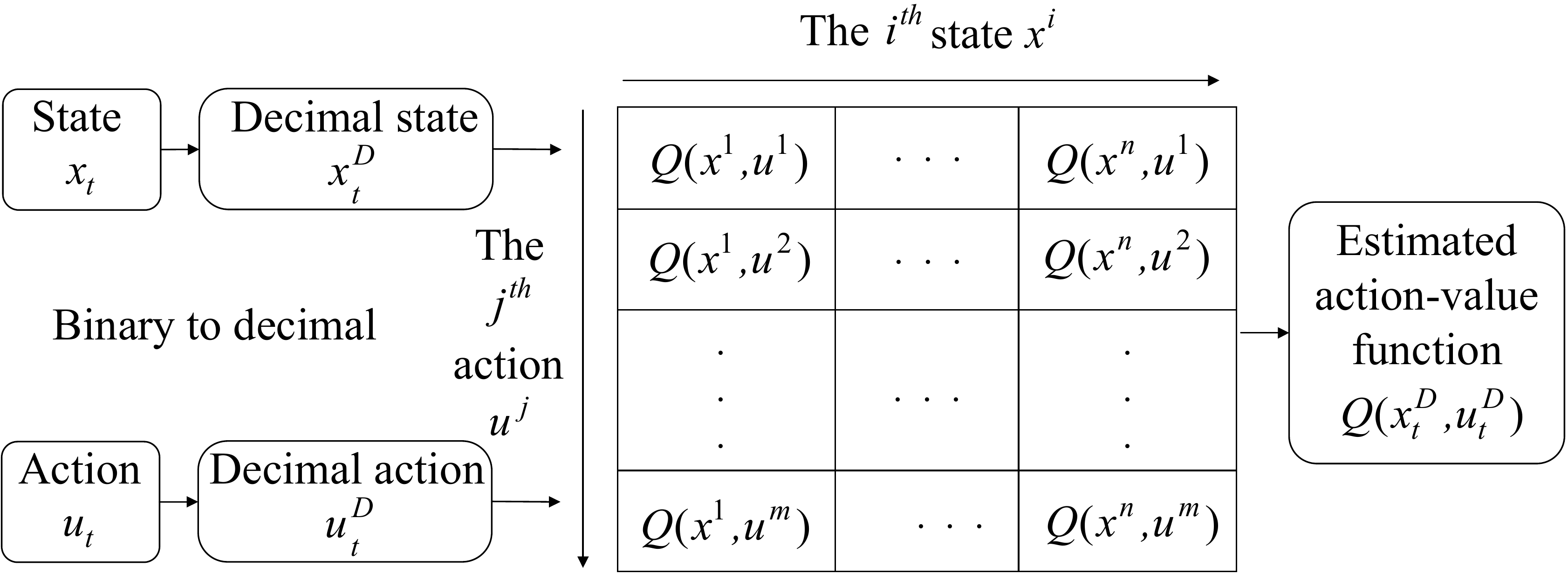}}
	\caption{$Q$-table for optimal infinite-horizon control of small-scale PBCNs}
	\label{figQtablePBCN}
\end{figure}
As shown in Figure \ref{figQtablePBCN}, to facilitate the representation of action-values in a $Q$-table, states and actions are converted from binary to decimal. Before conversion, the vector forms of a state $x_t\in \mathcal{B}^{n} $ and an action $u_t \in \mathcal{B}^{m} $ are inconvenient for an action-value $ Q(x_{t},u_{t})$ to be described in tabular form, so the conversion is considered. Convert $x_t$ from binary to decimal:
\begin{equation}
	x_t^{D}=\sum\limits_{j=1}^n2^{n-j}x_t(j),
\end{equation}
where $x_t(j),j=1,\dots,n$ is the $j^{th}$ component of $x_t$. $x_t^{D}\in\mathbf{X^{D}}$ is the decimal state, where $\mathbf{X^{D}}=\{0,1,\dots,2^n-1\}$. Similarly, convert $u_t$ from binary to decimal:
\begin{equation}
	u_t^{D}=\sum\limits_{j=1}^m2^{m-j}u_t(j),
\end{equation}
where $u_t(j),j=1,\dots,m$ is the $j^{th}$ component of $u_t$. $u_t^{D}\in\mathbf{U^{D}}$ is the decimal action, where $\mathbf{U^{D}}=\{0,1,\dots,2^m-1\}$. After conversion, we can easily find $Q(x_t^{D},u_t^{D})$ in the $u_t^{Dth}$ row and $x_t^{Dth}$ column of the $Q$-table. In the following, Algorithm 1, namely, optimal infinity-horizon control of small-scale PBCNs using $Q$L, is given.

\begin{algorithm}[htb] 
	\caption*{ $\mathbf{Algorithm\ 1}$ Optimal infinite-horizon control of small-scale PBCNs using $Q$L} 
	\label{alg:Framwork} 
	\begin{algorithmic}[1] 
		\REQUIRE ~~\\ 
		Learning rate $\alpha_{t'}\in (0,1]$, discount factor $\gamma\in [0,1) $, greedy rate $\epsilon\in[0,1]$, maximum of episodes $N$, maximum of time steps $T$\\
		\ENSURE ~~\\ 
		Optimal control policy $\pi^*(x_t^{D}),\forall x_t^{D}\in\mathbf{X^{D}}$
		\STATE Initialize $Q(x_t^{D},u_t^{D}) \leftarrow 0,\forall x_t^{D}\in \mathbf{X^{D}} ,\forall u_t^{D}\in \mathbf{U^{D}}$
		\FOR{ $ep = 0, 1,\dots, N-1 $ }
		\STATE $x_0^{D} \leftarrow \operatorname{rand}(\mathbf{X^{D}})$
		\FOR{$t=0,1,\dots, T-1 $}
		\STATE $u_{t}\leftarrow\left\{
		\begin{aligned}
			&\arg \max _{u_t^{D} \in \mathbf{U^{D}}} Q(x_{t}^{D}, u_t^{D}),&&P=1-\epsilon\\
			&\operatorname{rand}(\mathbf{U^{D}}),&&P=\epsilon
		\end{aligned}\right. $
		\STATE$Q(x_t^{D},u_t^{D})\leftarrow\alpha_{t'}(r_{t+1}+\gamma\max\limits_{u_{t+1}^{D} \in \mathbf{U^{D}}}Q(x_{t+1}^{D}, u_{t+1}^{D}))+(1-\alpha_{t'})Q(x_t^{D},u_t^{D})$
		\ENDFOR 
		\ENDFOR 
		\RETURN $\pi^*(x_t^{D})\leftarrow\arg \max\limits_{u_t^{D}\in\mathbf{U^{D}}}Q(x_t^{D},u_t^{D}),\forall x_t^{D}\in  \mathbf{X^{D}}$ 
	\end{algorithmic}
\end{algorithm}

Algorithm 1 represents optimal infinite-horizon control of small-scale PBCNs using $Q$L. An optimal deterministic control policy $\pi^*(x_t^{D}),\forall x_t^{D}\in\mathbf{X^D}$ is obtained through the algorithm. It is worth noting that the optimal deterministic control policy $\pi^*(x_t^{D})$ is not necessarily unique, as an action-value $Q(x_t^{D},u_t^{D})$ may be maximized under multiple actions. Based on $\pi^*(x_t^{D})$, the deterministic state feedback controller $u_t(x_t^{D})=\pi^*(x_t^{D})$ is obtained.

In Algorithm 1, $ep$ means the number of episodes. ``Episode" is a reinforcement learning term. An episode means a period of interaction that has passed through $T$ time steps from any initial state $x_0^{D}$. For optimal infinite-horizon control of PBCNs, the introduction of episode enriches an agent's understanding of each state, so as to improve the control policy. Before introducing the episode, the agent interacts with the environment continuously from an initial state $x_0^{D}$. It is likely for the agent to experience only the adjacent states of $x_0^{D}$, but not other states. In this case, the control policy is limited.

The maximum of time steps $T$ should be adjusted according to the scale of a PBCN and the difficulty of the control goal. Specifically, $T$ increases with the difficulty of the control goal and the scale of a PBCN. If $T$ is too small, the agent cannot obtain a positive reward before the end of an episode, so the learning speed will be affected. If $T$ is too large, the agent will be in a situation similar to the no episode one, where the control policy is limited.

$Q$L has a convergence guarantee, which ensures $\pi^*(x_t^{D}), \forall x_t^{D}\in  \mathbf{X^{D}}$ obtained by Algorithm 1 approaches to the optimal one.

$\mathbf{Theorem\ 2}$\cite{QLcovergence}: $Q(x_t^{D},u_t^{D})$ converges to the fixed point $q^*(x_t^{D},u_t^{D})$ with probability one under the following conditions:
\begin{enumerate}
	\item $\sum_{t'=0}^{\infty} \alpha_{t'}=\infty$ and $\sum_{t'=0}^{\infty} \alpha_{t'}^{2}<\infty $, where $t'=ep\times T+t$ represents the global steps;
	\item $\operatorname{var}\left[r_{t}\right] $ is finite.
\end{enumerate}

$Q$L can effectively solve optimal infinite-horizon control of small-scale PBCNs, while DD$Q$N needs to be considered for large-scale ones. Specifically, the operation of $Q$L is based on a $Q$-table with $|\mathbf{X}|\times|\mathbf{U}| $ values. For PBCNs, $|\mathbf{X}|\times|\mathbf{U}|$
increases exponentially with the number of nodes and
control inputs. When the number of nodes in PBCNs is so large that $|\mathbf{X}|\times|\mathbf{U}| $ exceeds the computer memory, $Q$L is no longer applicable. We define these PBCNs as large-scale. For large-scale PBCNs, consider DD$Q$N. DD$Q$N uses function approximation that requires less memory than a $Q$-table to represent action-values, so it can effectively solve optimal infinite-horizon control of large-scale PBCNs.

\subsection{Optimal Control of Large-scale PBCNs Using DDQN}
In this section, $\pi^*$ for optimal infinite-horizon control of large-scale PBCNs is obtained by DD$Q$N. DD$Q$N has three advantages to the problem. Firstly, DD$Q$N is model-free, which solves the difficulty of modeling PBCNs. Secondly, DD$Q$N is suitable for problems with large state space, which makes it effective for large-scale PBCNs. Finally, DD$Q$N uses function approximation, which has strong generalization.
\begin{figure}[!t]
	\centerline{\includegraphics[width=\columnwidth]{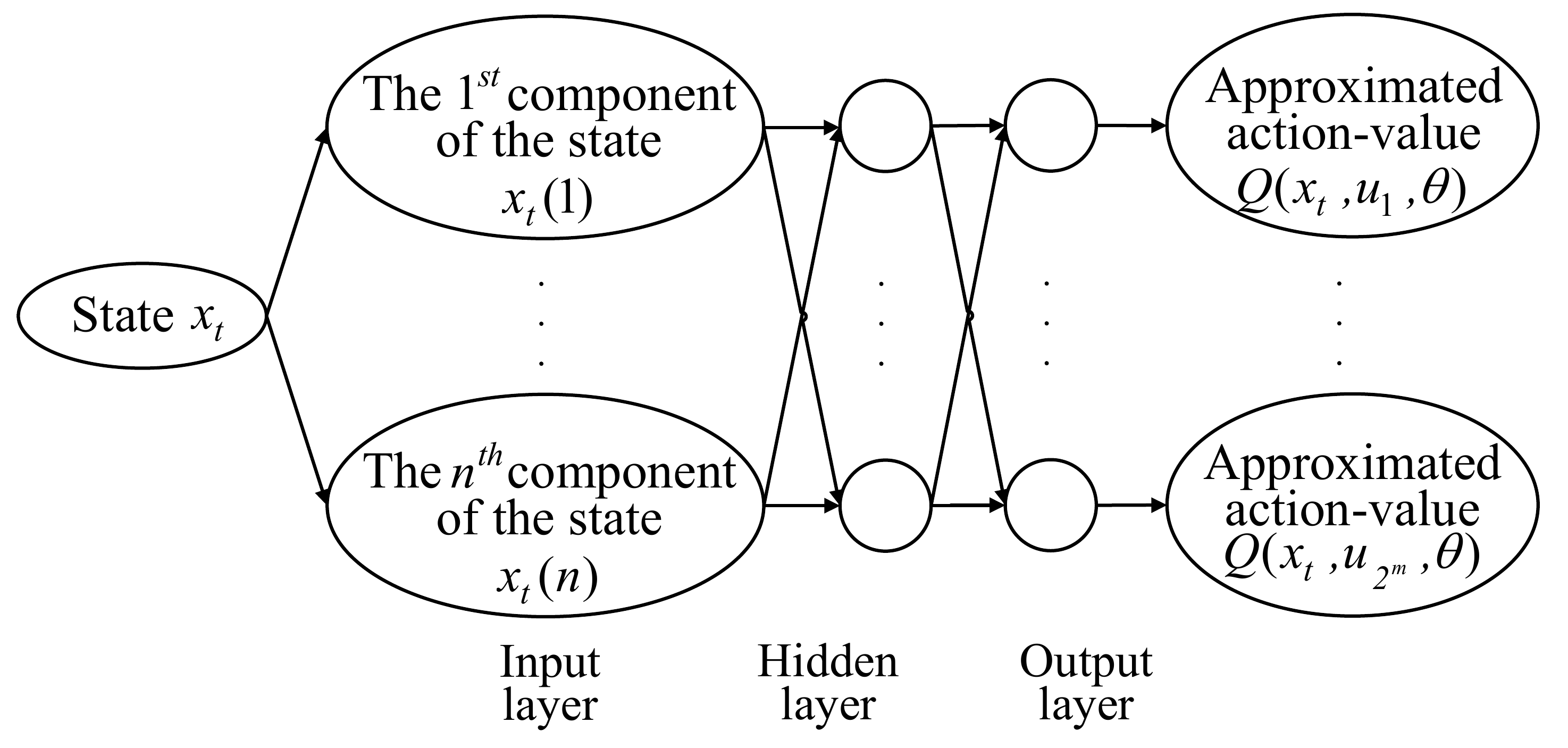}}
	\caption{Artificial neural network for optimal infinite-horizon control of large-scale PBCNs}
	\label{figAnnPBCN}
\end{figure}
\begin{algorithm}[htb] 
	\caption*{ $\mathbf{Algorithm\ 2}$ Optimal infinite-horizon control of large-scale PBCNs using DD$Q$N} 
	\begin{algorithmic}[1] 
		\REQUIRE ~~\\ 
		Learning rate $\beta\in (0,1]$, discount factor $\gamma\in[0,1) $, greedy rate $\epsilon\in[0,1]$, maximum of episodes $N$, maximum of time steps $T$, mini-batch size $\mathcal{M}$, target network learning rate $\tau\in[0,1]$, replay memory capacity $k$\\
		\ENSURE ~~\\ 
		Optimal control policy $\pi^*(x_t),\forall x_t\in  \mathbf{X}$
		\STATE Initialize the main network parameter $\theta\leftarrow \operatorname{rand}([0,1])$
		\STATE Initialize the target network parameter $\varphi\leftarrow \theta$
		\STATE Initialize the replay memory $\mathcal{D}_k \leftarrow\emptyset$
		\FOR{$ep = 0, 1,\dots, N-1 $}
		\STATE $x_0 \leftarrow \operatorname{rand}( \mathbf{X})$
		\FOR{$t=0,1,\dots, T-1 $}
		\STATE $u_{t}\leftarrow\left\{
		\begin{aligned}
			&\arg \max _{u_t \in \mathbf{U}} Q\left(x_{t}, u_t,\theta\right),&&P=1-\epsilon\\
			&\operatorname{rand}(\mathbf{U}),&&P=\epsilon
		\end{aligned}\right. $
		\STATE Take $u_t $, observe $x_{t+1} $ and $r_{t+1}$
		\STATE Store $(x_t,u_t,x_{t+1},r_{t+1}) $ at the rear of $\mathcal{D}_k$
		\WHILE{$|\mathcal{D}_k|\geq \mathcal{M}$}		
		\STATE Randomly select $\mathcal{M}$ sequences from $\mathcal{D}_k$ 
		\STATE for $i= 0, 1,\dots,\mathcal{M}-1$ do
		\STATE \quad $y_i \leftarrow r_i+\gamma Q(x_i',\arg\max\limits_{u\in\mathbf{U}}Q(x_i',u,\theta),\varphi) $
		\STATE end for
		\STATE $L(\theta)\leftarrow\frac{1}{\mathcal{M}}\sum\limits_i^\mathcal{M}(y_i-Q(x_i,u_i,\theta))^2$
		\STATE $\theta\leftarrow\theta-\alpha \nabla_\theta L(\theta)$
		\ENDWHILE
		\STATE $\varphi\leftarrow\beta\varphi+(1-\beta)\theta$
		\ENDFOR
		\ENDFOR 
		\RETURN $\pi^*(x_t)=\arg \max\limits_{u_t\in\mathbf{U}}Q(x_t,u_t,\theta),\forall x_t\in  \mathbf{X}$
	\end{algorithmic}
\end{algorithm}

In the following, Algorithm 2 is given, which represents optimal infinite-horizon control of large-scale PBCNs using DD$Q$N. An optimal deterministic control policy $\pi^*(x_t),\forall x_t\in \mathbf{X}$ is obtained through the algorithm. It is worth noting that $\pi^*$ is not necessarily unique, as an action-value $Q(x_t,u_t,\theta) $ may be maximized under multiple actions. Based on $\pi^*(x_t)$, the deterministic state feedback controller $u_t(x_t)=\pi^*(x_t)$ is obtained.

The artificial neural network for optimal infinite-horizon control of PBCNs is shown in Figure 6. The inputs of the artificial neural network is the components of a state $x_t(i),i=1,\dots,n$. This expression is a full and concise representation of the features of a state because reducing any one of the inputs will destroy the integrity, while increasing the inputs are not needed. Notice that the state $x_t$ need not be to be converted from binary to decimal, since $[x_t(1),...,x_t(n)]$ is a natural one-hot vector which is a suitable input for the artificial neural network\cite{geron2022hands}. 

The setting of the replay memory capacity $k$ and the mini-batch size $\mathcal{M}$ depends on the complexity of PBCN dynamics. Specifically, $k$ and $\mathcal{M}$ rise with the number of nodes $n$, the number of control inputs $m$, and the number of available logic function $l_i, i \in 1,\dots, n$. If  $k$ or $\mathcal{M}$ is too small, samples stored in replay memory and used for network training cannot well represent PBCN dynamics, so the control policy is likely to be limited. If $k$ or $\mathcal{M}$ is too large, the learning speed will be affected due to too much old-fashion experience.

$\mathbf{Remark\ 2:}$ In this paper, we do not prove the convergence of Algorithm 2, which is based on DD$Q$N. As two ANNs are used in calculating $y_i$, the behavior of DD$Q$N becomes very complex. Thus, the convergence of DD$Q$N is still an open problem, to the best of our knowledge.

\subsection{Computational Complexity}
DD$Q$N is a combination of $Q$L and deep learning. In terms of algorithm scalability, DD$Q$N can solve large-scale problems that $Q$L cannot. However, the time complexity of DD$Q$N is significantly higher than that of $Q$L.

Consider time complexity. In both $Q$L and DD$Q$N, an agent needs to select the control inputs with the optimal action-value from $2^m $ control inputs. The time complexity involved here is $O\left(2^{m}\right) $. In each episode of $N$ episodes, this operation takes $T$ steps. So, the total time complexity of this part is $O\left(N T 2^{m}\right)$. In addition, it is necessary to update network parameters in DD$Q$N, but not in $Q$L. It is known that the number of nodes in the input layer and output layer is $n$ and $2^m$ respectively. Assume that there is only 1 hidden layer with $h$ nodes. Then, the time complexity involved in this part is $O\left(2^{m}h+hn\right) $. This operation also takes $T$ steps in $N$ episodes, where $NT$ should rise with the growth of the state number $2^n$ and the complexity of the dynamics of the PBCNs\cite{QLStabilityPBCN2021}. So, the total time complexity of this part is $O\left(N T 2^{m}h+NThn\right) $. In a word, the additional operation of network updating makes the time complexity of DD$Q$N higher than $Q$L.

Consider space complexity. The space complexity of $Q$L depends on the size of the $Q$-table. When the number of control inputs is $m$ and the number of nodes is $n$, the space complexity involved in this part is $O\left(2^{n+m}\right) $. The space complexity of DD$Q$N depends on the size of network parameters $\theta$ and $\varphi$, and replay memory $\mathcal{D}_k $. It is known that the number of nodes in an input layer and an output layer is $n$ and $2^m$, respectively. Assume that there is only 1 hidden layer with $h$ nodes. Then, the space complexity required by storing network parameters $\theta$ and $\varphi$ is $O\left(2^{m+1}h+2hn\right) $. To break the correlation between samples, the capacity of replay memory must be large enough. This increases the space complexity of DD$Q$N significantly. Whether the space complexity of $Q$L or DD$Q$N  is lower depends on the specific problem and parameter settings. Generally speaking, the space complexity of $Q$L is lower for small-scale PBCNs, while the space complexity of DD$Q$N is lower for large-scale PBCNs.

To sum up, $Q$L is a better choice compared with DD$Q$N in terms of time complexity. The advantage of DD$Q$N is that it can solve optimal infinite-horizon control of large-scale PBCNs, while $Q$L or model-based methods cannot. The next section uses an example to illustrate the applicability of DD$Q$N in large-scale PBCNs. Specifically, DD$Q$N is successfully applied to optimal infinite-horizon control of the PBCN with 28 nodes and 3 control inputs, whose number of state-action pairs is $2.15\times10^9 $.

\section{Simulation}
In this section, the performances of $Q$L and DD$Q$N for optimal infinite-horizon control of PBCNs are analyzed and compared based on simulation. We consider two examples, which are a small-scale PBCN with 3 nodes and 1 control input, and a large-scale PBCN with 28 nodes and 3 control inputs. For each example, the change of the rewards in the training process and the obtained optimal controller are shown. For the small-scale PBCN, we analyze the optimal action-value errors and the optimal control policy errors in both $Q$L and DD$Q$N. From the errors, the convergence of $Q$L and DD$Q$N is shown.

\subsection{Examples}
To evaluate the performance of the proposed algorithms, optimal action-value error $ErrorQ_{ep}$ and optimal control policy error $Error\pi_{ep}$ are defined. The optimal action-value error $ErrorQ_{ep}$ is the average absolute difference of optimal action-values between $Q$L or DD$Q$N and policy iteration at the end of the episode $ep$:
\begin{equation}
	ErrorQ_{ep} =
	\frac{1}{|\mathbf{X}|}\sum\limits_{\forall x\in\mathbf{X}}|v^*_{PI}(x)-\max\limits_{u\in\mathbf{U}}Q_{ep}(x,u(,\theta))|,
\end{equation}
where $\max\limits_{u\in\mathbf{U}}Q_{ep}(x,u(,\theta)) $ is the optimal action-value of the state $x$ based on $Q$L or DD$Q$N at the end of the episode $ep$, and $v^*_{PI}(x) $ is the optimal action-value of the state $x$ obtained by policy iteration. Policy iteration is a model-based algorithm that converges to the optimal state-values and the optimal policy in finite-time \cite{PI2018Wu}. Thus, it is reasonable to compare the optimal action-values and the optimal control policy obtained by $Q$L or DD$Q$N with the ones obtained by policy iteration. Similarly, optimal control policy error $Error\pi_{ep}$ is the average absolute difference of the optimal control policy between $Q$L or DD$Q$N and policy iteration at the end of the episode $ep$:
\begin{equation}
	Error\pi_{ep} =
	\frac{1}{|\mathbf{X}|}\sum\limits_{\forall x\in\mathbf{X}}||\pi^*_{PI}(x)-\pi^*_{ep}(x)||,
\end{equation}
where $\pi^*_{PI}(x) $ is the optimal action at the state $x$ obtained by policy iteration, and $\pi^*_{ep}(x)$ is the one obtained by $Q$L or DD$Q$N at the end of the episode $ep$. Notice that an optimal action of a PBCN is expressed as a vector $\pi^*(x):=\left(\pi_{1}^*(x), \ldots, \pi_{m}^*(x) \right) \in \mathcal{B}^{m} $, which contains $m$ elements. The distance between $\pi^*_{PI}(x) $ and $\pi^*_{ep}(x)$ under the norm $||.|| $ is the average difference between their components:
\begin{equation}
	||\pi^*_{PI}(x)-\pi_{ep}^*(x)||=\frac{1}{m}\sum\limits_{i=1}^m|\pi^*_{PIi}(x)-\pi^*_{epi}(x)|,
\end{equation}
where $\pi^*_{epi}(x) $ is the $i^{th}$ component of $\pi^*_{ep}(x)$, and $\pi^*_{PIi}(x) $ is the $i^{th}$ component of $\pi^*_{PI}(x)$.

In the following examples, our control goals can be divided into two parts, which are to make control inputs and nodes in the desired form. According to the two-part goal, $l_{i+1}$ is expressed as follows:
\begin{equation}
	l_{t+1}(x_t,u_t) = W_pP_{t+1}+W_hH_{t+1}.
	\label{eqr}
\end{equation}
In (\ref{eqr}), $P_{t+1}=[p_{t+1}^1(u_t(1)),p_{t+1}^2(u_t(2)),\dots,p_{t+1}^m(u_t(m))]^T$ is the control input cost, where $p_{t+1}^i(u_t(i)):\mathcal{B}\rightarrow\mathbb{R}^{+}$ is the cost of the $i^{th}$ control input. $H_{t+1}=[h_{t+1}^1(x_t(1)), h_{t+1}^2(x_t(2)),\dots,  h_{t+1}^n(x_t(n))]^T$ is the state cost, where $h_{t+1}^i(x_t(i)):\mathcal{B}\rightarrow\mathbb{R}^{+}$ is the cost of the $i^{th}$ node. $W_p=[w_{p1},w_{p2},\dots,w_{pm}]$ is the weight of the control input cost, where $w_{pi}$ is the weight of $p_{t+1}^i(u_t(i))$. Besides, $W_h=[w_{h1},w_{h2},\dots,w_{hn}]$ is the weight of the state cost, where $w_{hi}$ is the cost of $h_{t+1}^i(x_t(i))$. If one of our goals is to make $u_t(i),\forall t \in \mathbb{Z^{+}}$ in a specific form $u_i^*$, then $p_{t+1}^i(u_t(i))$ is defined as follows:
\begin{equation}
	p_{t+1}^i(u_t(i))=
	\left\{
	\begin{aligned}
		&0,&&\text{if}\ u_t(i)=u_i^*.\\
		&1,  &&\text{else}.
	\end{aligned}
	\right.
\end{equation}
Similarly, if another goal is to make $x_t(i), \forall t \in \mathbb{Z^{+}}$ in a specific form $x_i^*$, then $h_{t+1}^i(x_t(i))$ is defined as follows:
\begin{equation}
	h_{t+1}^i(x_t(i))=
	\left\{
	\begin{aligned}
		&0,&&\text{if}\ x_t(i)=x_i^*.\\
		&1,  &&\text{else}.
	\end{aligned}
	\right.
\end{equation}
If the $i^{th}$ control input or the $i^{th}$ node has no impact on the goal, then set $p_{t+1}^i(u_t(i))=0$ or $h_{t+1}^i(x_t(i))=0$. Besides, $w_{pi}$ or $w_{hi}$ depends on the importance of the $i^{th}$ control input or the $i^{th}$ node to the goal.

$\mathbf{Example\ 1:}$ We consider the PBCN model of the apoptosis network given in [28]. The PBCN has 3 nodes $x_t=\left(x_{t}(1), x_{t}(2), x_{t}(3) \right) \in \mathcal{B}^{3} $ and 1 control input $u_t=\left(u_{t}(1) \right) \in \mathcal{B}^{1} $. The PBCN dynamics are given as follows:
\begin{equation}
	\left\{\begin{aligned}
		x_{t+1}(1)=f_{1}\left(x_{t}(1), x_{t}(2), x_{t}(3), u_{t}(1)\right), \\
		x_{t+1}(2)=f_{2}\left(x_{t}(1), x_{t}(2), x_{t}(3), u_{t}(1)\right), \\
		x_{t+1}(3)=f_{3}\left(x_{t}(1), x_{t}(2), x_{t}(3), u_{t}(1)\right),
	\end{aligned}\right.   \ t\in \mathbb{Z^{+}},
	\label{eqPBCNsmall}
\end{equation}
where the logic functions are represented as follows:
\begin{equation}
	\left\{\begin{aligned}
		&f_{1}^{1}\left(x_{t}(1), x_{t}(2), x_{t}(3), u_{t}(1)\right)=\neg x_{t}(2) \wedge u_{t}(1), 
		&P_1^1=0.6, \\
		&f_{1}^{2}\left(x_{t}(1), x_{t}(2), x_{t}(3), u_{t}(1)\right)= u_{t}(1),
		&P_1^2=0.4,\\
		&f_{2}^{1}\left(x_{t}(1), x_{t}(2), x_{t}(3), u_{t}(1)\right)= \neg x_{t}(1) \wedge x_{t}(3), 
		&P_2^1=0.7,\\
		&f_{2}^{2}\left(x_{t}(1), x_{t}(2), x_{t}(3), u_{t}(1)\right)=x_{t}(2),
		&P_2^2=0.3,\\
		&f_{3}^{1}\left(x_{t}(1), x_{t}(2), x_{t}(3), u_{t}(1)\right)= x_{t}(2) \vee u_t(1),
		&P_3^1=0.8,\\
		&f_{3}^{2}\left(x_{t}(1), x_{t}(2), x_{t}(3), u_{t}(1)\right)= x_{t}(3), 
		&P_3^2=0.2.
	\end{aligned}\right. 
\end{equation}

Consider the optimal infinite-horizon control of the PBCN (\ref{eqPBCNsmall}). We aim to find $\tilde{\pi}^*$ which minimizes the cost-to-go function: 
\begin{equation}
	J_{\tilde{\pi}}(x_t)=\lim_{N\rightarrow \infty}\mathbb{E}_{\tilde{\pi}}\left[\sum_{i=t}^{N} \gamma ^{i-t}l_{i+1}|x_t\right], \forall t\in\mathbb{Z^{+}},
\end{equation}
where the cost $l_{i+1}\left(x_t,u_t\right)$ is related to the control goals. One of the goals is to increase the activity of $x_t(2)$, i.e., let $x_t(2)=1, \forall t \in \mathbb{Z^{+}}$ if possible. The goal is expressed as the state cost:
\begin{equation}
	h_{t+1}^2(x_t(2))=\left\{
	\begin{aligned}
		&0,&\text{if}\ x_t(2)=1.\\
		&1,  &\text{if}\ x_t(2)=0.
	\end{aligned}\right.
\end{equation}
Another goal is to make the control input $u_t(1)=0, \forall t \in \mathbb{Z^{+}}$ if possible. This goal is expressed as the control input cost:
\begin{equation}
	p_{t+1}^1(u_t(1))=\left\{\begin{aligned}
		&0,&\text{if}\ u_t(1)=0.\\
		&1,  &\text{if}\ u_t(1)=1.
	\end{aligned}\right.
\end{equation}
Since $x_t(1)$ and $x_t(3)$ has nothing to do with the goal, set $h_{t+1}^1(x_t(1))=0$ and $h_{t+1}^3(x_t(3))=0$. Then, we define $w_{h2}=0.8$ and $w_{p1}=0.2$. According to (\ref{eqr}), the cost is obtained as follows:
\begin{equation}
	l_{t+1}(x_t,u_t)=\left\{\begin{aligned}
		&0,  &\text{if}\ u_t(1)=0\ \text{and}\ x_t(2)=1.\\
		&0.2,&\text{if}\ u_t(1)=1\ \text{and}\ x_t(2)=1.\\
		&0.8,&\text{if}\ u_t(1)=0\ \text{and}\ x_t(2)=0.\\
		&1,  &\text{if}\ u_t(1)=1\ \text{and}\ x_t(2)=0.\\
	\end{aligned}\right.
\end{equation}

We turn the problem into finding $\pi^*$ which maximizes the action-value function:
\begin{equation}
	q_\pi(x_t,u_t)= \mathbb{E}_\pi\left[\sum_{i=t}^{\infty} \gamma ^{i-t}r_{i+1}|x_t,u_t\right],\forall x_t\in \mathbf{X},
\end{equation}
where $r_{i+1}$ is defined as $r_{i+1}=-l_{i+1}+1$ to meet the condition of Theorem 1. The specific version of $r_{i+1}$ is given as follows:
\begin{equation}
	r_{t+1}(x_t,u_t)=\left\{\begin{aligned}
		&1,  &\text{if}\ u_t(1)=0\ \text{and}\ x_t(2)=1,\\
		&0.8,&\text{if}\ u_t(1)=1\ \text{and}\ x_t(2)=1,\\
		&0.2,&\text{if}\ u_t(1)=0\ \text{and}\ x_t(2)=0,\\
		&0,  &\text{if}\ u_t(1)=1\ \text{and}\ x_t(2)=0.\\
	\end{aligned}\right.
\end{equation}
Following the above definition, we can conclude that $\pi^*$ is equivalent to $\tilde{\pi}^*$. Then, Algorithms 1 and 2 are used to obtain $\pi^*$ for optimal infinite-horizon control of the PBCN. 
\begin{figure}[!t]
	\centerline{\includegraphics[width=\columnwidth]{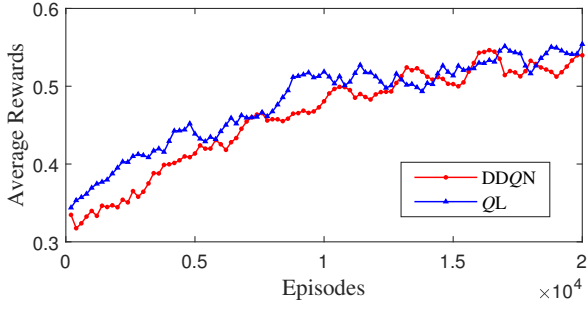}}
	\caption{Average rewards during training}
	\label{figSTrain}
\end{figure}

The average rewards in the training process by DD$Q$N and $Q$L are shown in Figure \ref{figSTrain}. The word ``average” means taking an average of rewards in the neighboring 1000 episodes. The reason for taking the average is to reduce the influence of initial states on rewards, and then show the influence of training on rewards more objectively. The average rewards of both DD$Q$N and $Q$L keep rising from 0.35 to 0.55. For the training time, DD$Q$N takes 2 hours, while $Q$L only takes 10 seconds. From the above results and the time complexity, it can be concluded that $Q$L is less time-consuming than DD$Q$N in the optimal infinite-horizon control of PBCNs.

\begin{figure}[!t]
	\centerline{\includegraphics[width=\columnwidth]{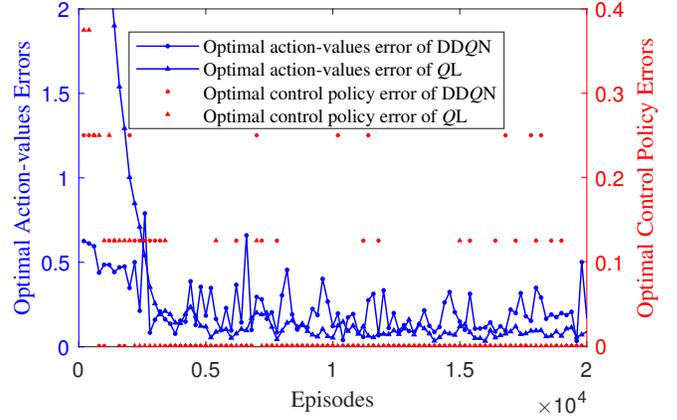}}
	\caption{Errors during training}
	\label{figSError}
\end{figure}

The convergence of DD$Q$N and $Q$L in the PBCN is shown in Figure \ref{figSError}. The variation of the optimal action-value errors $ErrorQ_{ep}$ and the optimal control policy errors $Error\pi_{ep}$ with the number of episodes are described. No matter DD$Q$N or $Q$L, the errors decrease with the increase of episode number. In the short run, the optimal control policy obtained from DD$Q$N is more likely to approach the fixed point, while in the long run, the one obtained from $Q$L is more likely to approach the fixed point. Therefore, for optimal infinite-horizon control of small-scale PBCNs, $Q$L is more recommended if the number of training steps can be large enough. It is worth mentioning that at the end of the episode, the optimal control policy obtained from both DD$Q$N and $Q$L converges to the fixed point, which is:
\begin{equation}
	\pi^*(x_t)=\left\{
	\begin{aligned}
		&1, &&\text{if}\ x_t=(0,0,0)\ \text{or}\  x_t=(1,0,0).\\
		&0, &&\text{others}.
	\end{aligned}\right.  
	\label{eqpolicys}
\end{equation}

$\mathbf{Remark\ 3:}$ The policy (\ref{eqpolicys}) obtained by DD$Q$N and $Q$L is exactly the same as the one according to policy iteration \cite{PI2006Pal,PI2018Wu,PI2019WU,PI2021Wu,STP2014Fornasini}, which shows the optimality of (\ref{eqpolicys}).

\begin{figure}[!t]
	\centerline{\includegraphics[width=\columnwidth]{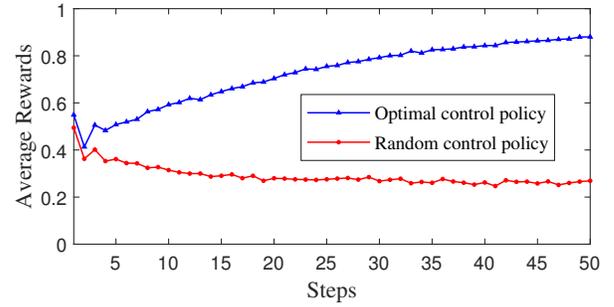}}
	\caption{Average rewards under optimal and random policy}
	\label{figSCR}
\end{figure}
\begin{figure}[!t]
	\centerline{\includegraphics[width=\columnwidth]{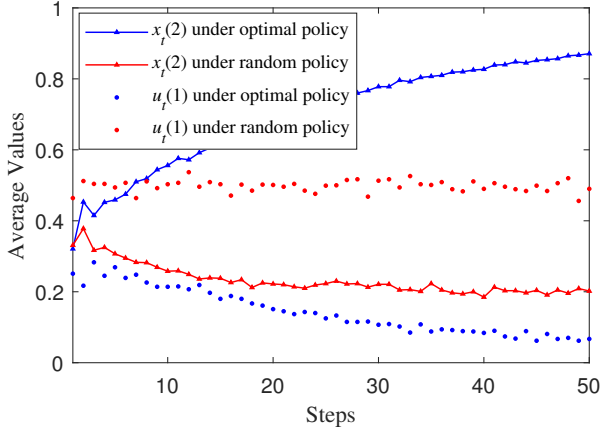}}
	\caption{Average values under optimal and random policy}
	\label{figSCV}
\end{figure}
The effect of optimal controller obtained is shown in Figures \ref{figSCR} and \ref{figSCV}. Figure \ref{figSCR} describes the average rewards according to the optimal control policy, and presents the ones according to the random control policy as a comparison. The word ``average" means taking an average of rewards in 1000 repeated experiments with random initial states. The reason for taking an average is to reduce the influence of initial states on rewards, so as to describe the effect of optimal control policy on rewards more objectively. As shown in Figure \ref{figSCR}, the average rewards according to the optimal control policy are significantly greater than that according to the random control policy. In particular, according to the optimal control policy, the average rewards increase from 0.4 to 0.9 with the number of steps, which are close to the maximum reward 1. In contrast, according to the random control policy, the average rewards fluctuate around 0.3, far from the maximum reward 1. Figure \ref{figSCV} describes the average value of the target gene activity $x_t(2)$ and the target control input $u_t(1)$ according to the optimal control policy, and the ones according to the random policy control as a comparison. Similarly, the word ``average" means taking an average of rewards in 1000 repeated experiments with random initial states. As shown in Figure \ref{figSCV}, the average values of $x_t(2)$ and $u_t(1)$ according to the optimal control policy are closer to the control goal than that according to the random control policy. To be specific, according to the optimal control policy, $x_t(2)$ increases from 0.3 to 0.9 with the number of steps, which is close to the control goal 1, and $u_t(1)$ decreases from 0.2 to 0.05, which is close to the control goal 0. In contrast, according to the random control policy, $x_t(2)$ decreases from 0.3 to 0.2, while $u_t(1)$ fluctuates around 0.5, far from the control goal.

$\mathbf{Remark\ 4:}$ In our paper, the system model is presented only for illustrating the problem, but the agent has no knowledge of it. 

$\mathbf{Example\ 2:}$ We consider the PBCN model of the reduced-order T-cell given in [32]. The PBCN has 28 nodes $x_t=\left(x_{t}(1), \dots, x_{t}(28) \right) \in \mathcal{B}^{28} $ and 3 control inputs $u_t=\left(u_{t}(1),u_{t}(2),u_{t}(3) \right) \in \mathcal{B}^{3} $. The PBCN dynamics are abbreviated as follows:
\begin{equation}
	\begin{aligned}
		&x_1^{+}=x_6\wedge x_{13};\ x_2^{+}=x_{25};\ x_3^+=x_2;\ x_4^+=x_{28};\ x_5^+=x_{21};\ x_6^{+}=\\
		&x_5;\ x_7^+=(x_{15}\wedge u_2)\vee(x_{26}\wedge u_2);\ x_8^+=x_{14};\ x_9^+=x_{18};\ x_{10}^+=\\ 
		&x_{25}\wedge x_{28};\ x_{11}^+=\neg x_9;\ x_{12}^+=x_{24};\  x_{13}^+=x_{12};\ x_{14}^+=x_{28};\ x_{15}^+=\\
		&(\neg x_{20})u_1\wedge u_2;\ x_{16}^+=x_3;\ x_{17}^+=\neg x_{11};\ x_{18}^+=x_2;\ x_{19}^+=(x_{10}\wedge\\
		& x_{11}\wedge x_{25}\wedge x_{28})\vee(x_{11}\wedge x_{23}\wedge x_{25}\wedge x_{28});\ x_{20}^+=x_7\vee \neg x_{26};\ \\
		&x_{21}^+=x_{11}\vee x_{22};\ x_{22}^+=x_2\wedge x_{18};\ x_{23}^+=x_{15};\ x_{24}^+=x_{18};\ x_{25}^+=\\
		&x_8;\ x_{26}=\neg x_4\wedge u_3,\ \text{P}=0.5\ x_{26}=x_{26},\ \text{P}=0.5;\ x_{27}^+=x_7\vee\\
		&(x_{15}\wedge x_{26});\ x_{28}^+=\neg x_4\wedge x_{15}\wedge x_{24},
	\end{aligned}
\label{eqPBCNB}
\end{equation}
where $x_i$ represents the activity of the $i^{th}$ gene at the current time step, and $x_i^+$ represents the activity of the $i^{th}$ gene at the next time step.

We consider the optimal infinite-horizon control of the PBCN (\ref{eqPBCNB}). We aim to find $\tilde{\pi}^*$ which minimizes the cost-to-go function in (\ref{eqcosttogo}): 
\begin{equation}
	J_{\tilde{\pi}}(x_t)=\lim_{N\rightarrow \infty}\mathbb{E}_{\tilde{\pi}}\left[\sum_{i=t}^{N} \gamma ^{i-t}l_{i+1}|x_t\right], \forall t\in\mathbb{Z^{+}},
\end{equation}
where $l_{i+1}\left(x_t,u_t\right)$ is related to the control goal. One of the control goals is to decrease the activity of $x_t(1)$ and $x_t(7)$, i.e., let $x_t(1)=0$ and $x_t(7)=0$. The goal is expressed as the state cost, which is given as follows:
\begin{equation}
	h_{t+1}^i(x_t(i))=\left\{
	\begin{aligned}
		0, &\ \text{if}\ x_t(i)=0,\\
		1, &\ \text{if}\ x_t(i)=1,\\
	\end{aligned}\right. \ i\in\{1,7\}.
\end{equation}
Another goal is to make control inputs be the ones $u_t(1)=0$, $u_t(2)=0$, and $u_t(3)=0$ if possible. The goal is expressed as the control input cost, which is given as follows:
\begin{equation}
	p_{t+1}^i(u_t(i))=
	\left\{
	\begin{aligned}
		0, &\ \text{if}\  u_t(i)=0,\\
		1, &\ \text{if}\  u_t(i)=1,\\
	\end{aligned}\right. \ i\in\{1,2,3\}.
\end{equation}
In addition to the mentioned nodes, others do not affect the goal. So, $h_{t+1}^i(x_t(i))=0,\ i\not=1 \ \text{and}\ i\not=7$. Then, we define $w_{h1}=0.4$, $w_{h7}=0.3$,  $w_{p1}=0.1$, $w_{p2}=0.1$, and $w_{p3}=0.1$. The cost is obtained according to (\ref{eqr}). 

We define $r_{i+1}=-l_{i+1}+1, \forall i \in \mathbb{Z}^{+}$. Then, according to Theorem 1, the problem is turned into finding $\pi^*$ which maximizes the action-value function:
\begin{equation}
	q_\pi(x_t,u_t)= \mathbb{E}_\pi\left[\sum_{i=t}^{\infty} \gamma ^{i-t}r_{i+1}|x_t,u_t\right],\forall x_t\in \mathbf{X}.
\end{equation}

The PBCN (\ref{eqPBCNB}) has 28 nodes and 3 control inputs. The corresponding number of the action-values is $2.15\times10^9$, which is so large that $Q$L or model-based methods like policy iteration are no longer applicable. For optimal infinite-horizon control of large-scale PBCNs, DD$Q$N is used to obtain optimal control policies.
\begin{figure}[!t]
	\centerline{\includegraphics[width=\columnwidth]{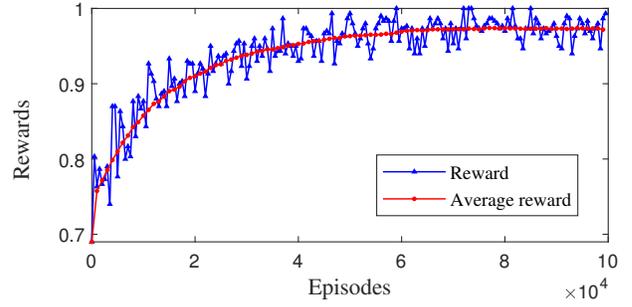}}
		\caption{Rewards during training}
	\label{figBTrain}
\end{figure}

The rewards and the average rewards in the training process of DD$Q$N are described in Figure \ref {figBTrain}. The word ``average” means taking an average of rewards in the neighboring 1000 episodes. As the number of episodes increases, both the rewards and the average rewards keep rising, close to the maximum reward 1. The average rewards increase fast in the first 60000 episodes, and almost stop rising and converge in the later 40000 episodes. The growth trend of the rewards is basically consistent with the one of the average rewards. Taking the average rewards as the reference, the rewards float with $\pm1$ as the episodes change, which is mainly due to the difference in initial states. As the scale of a PBCN increases, the maximum number of episodes and time steps also increase, which leads to a longer training time, i.e., 56 hours.
\begin{figure}[!t]
	\centerline{\includegraphics[width=\columnwidth]{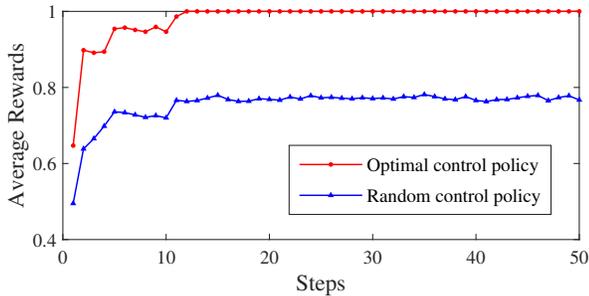}}
	\caption{Average rewards under optimal and random policy}
	\label{figBCR}
\end{figure}
\begin{figure}[!t]
	\centerline{\includegraphics[width=\columnwidth]{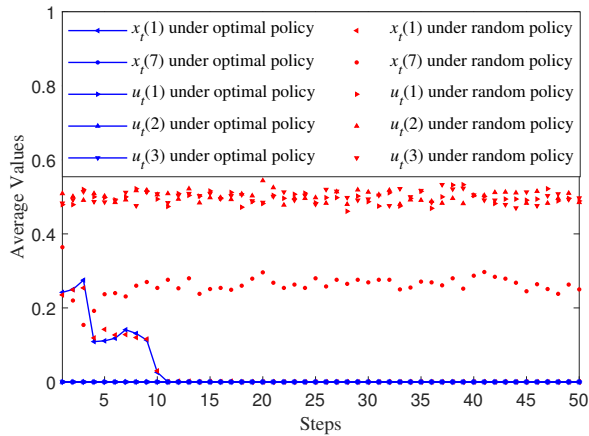}}
	\caption{Average values under optimal and random policy}
	\label{figBCV}
\end{figure}

The effect of the optimal controller obtained by DD$Q$N is shown in Figures \ref{figBCR} and \ref{figBCV}. Figure \ref{figBCR} describes the average rewards according to the optimal control policy, and presents the ones according to the random control policy as a comparison. The word ``average" means taking an average of rewards in 1000 repeated experiments with random initial states. As shown in Figure \ref{figBCR}, the average rewards according to the optimal control policy are significantly greater than that according to the random control policy. In particular, according to the optimal control policy, the average rewards increase from 0.9 to the maximum reward 1 with the number of steps. In contrast, according to the random control policy, the average rewards increase from 0.65 to 0.8, far from the maximum reward 1. Figure \ref{figBCV} describes the average value of the target gene activity $x_t(1)$ and $x_t(7)$ and the control inputs $u_t(1)$, $u_t(2)$, and $u_t(3)$ according to the optimal policy control, and the ones according to the random policy control as a comparison. As shown in Figure \ref{figBCV}, the average values of $x_t(1)$, $x_t(7)$, $u_t(1)$, $u_t(2)$, and $u_t(3)$ according to the optimal control policy are close to the control goal. To be specific, according to the optimal control policy, $x_t(1)$, $x_t(7)$, $u_t(1)$, $u_t(2)$, and $u_t(3)$ all decrease to the control goal 0 at the time step 11. In contrast, according to the random control policy, $x_t(7)$ fluctuates around 0.25, and all the control inputs fluctuate around 0.5, far from the control goal.

From the above results, it is concluded that DD$Q$N can solve the infinite-horizon optimal control of large-scale PBCNs, thus having certain advantages over the methods in previous literatures. Model-based optimal control mostly uses semi-tensor product and policy iteration \cite{PI2006Pal,PI2018Wu,PI2019WU,PI2021Wu,STP2014Fornasini}. These methods are based on matrix operation, so they are not suitable for large-scale PBCNs, generally speaking, the ones with more than 20 nodes. In contrast, we propose DD$Q$N, which can handle large-scale PBCNs. Besides, the sample-based method using path integral given by \cite{KLsamplebased} has two advantages, which are model-free and applicable to large-scale PBCNs. However, the method can only give the optimal policy according to a given initial state. Compared with \cite{KLsamplebased}, the optimal policy can be obtained without knowledge of initial states in our method, due to the strong generalization of DD$Q$N.

\subsection{Pattern and Details}
The simulation was completed on a 6-Core Intel i5-6200U processor with a frequency of 2.30GHz, and 12GB RAM. The software which we used is MATLAB R2021a. $Q$L and policy iteration were implemented by scripts, and DD$Q$N was implemented by the Reinforcement Learning Designer Toolbox.
\begin{table}
	\caption{Parameter settings}
	\label{table}
	\setlength{\tabcolsep}{2.5pt}
	\begin{center}
	\begin{tabular}{|p{38pt} p{30pt} p{30pt}  p{30pt} p{30pt} p{30pt} p{30pt}|}
		\hline
		Example & $N$           & $T$ & $\mathcal{M}$ & $k$           &   $h$ 
		& $\delta $
		\\
		\hline
		1       &$2\times 10^4$ & $15$&
		128           & $5\times10^4$ &   $2$
		&  $8\times 10^{-6} $\\
		2       &$1\times 10^5$ & $30$&
		256           & $2\times10^5$ &   $16$
		&  $2\times 10^{-6} $\\
		\hline
	\end{tabular}
	\end{center}
	\label{tabDetail}
\end{table}

In terms of parameter selection, for both Examples 1 and 2, we chose the same discount factor $\gamma=0.9 $, target network learning rate $\beta=0.001 $, the number of hidden layers $n_h=1$, and rectified linear units (ReLU) as activation function. In Example 1, we set the learning rate for $Q$L as a generalized harmonic series $a_{t'}=1/(ep+1)^\omega$ where $\omega=0.6$, which satisfies conditions 1) and 2) in Theorem 2. According to the complexity of Examples 1 and 2, we selected different maximum number of episodes $N$, maximum number of time step $T$, mini-batch size $\mathcal{M}$, replay memory capacity $k$, the number of nodes in a hidden layer $h$, and decay rate of greedy rate $\delta $. The specific parameter settings are shown in Table \ref {tabDetail}. Besides, we defined greedy rate as $\epsilon=(1-\delta) ^{t'}$.

\section{Conclusion}
In this paper, a deep reinforcement learning based method is proposed to obtain optimal infinite-horizon control policies for PBCNs. In particular, we establish the connection between action-value functions in deep reinforcement learning and cost-to-go functions in traditional optimal control. Then, $Q$L and DD$Q$N are applied to optimal infinite-horizon control of small-scale PBCNs and large-scale PBCNs, respectively. Meanwhile, the optimal state feedback controllers are designed. The proposed method in this paper has two advantages. First, both $Q$L and DD$Q$N are model-free techniques, which resolve the difficulty of modeling PBCNs. Second, DD$Q$N can solve optimal infinite-horizon control of large-scale PBCNs while many algorithms cannot. Through the discussion of computational complexity and the simulation, we compare the advantages of $Q$L and DD$Q$N. The advantage of $Q$L over DD$Q$N lies in its convergence guarantee and small time complexity. Therefore, optimal control policies can be obtained through $Q$L with higher accuracy in a shorter time. However, the premise of $Q$L is that the memory required for action-values is within the RAM of the computer. Therefore, we defined the small-scale PBCNs to which $Q$L is applicable as those that meet the memory requirement. And for others, we recommend using DD$Q$N, which can solve problems with large state space. It is worth mentioning that although the convergence of DD$Q$N has not been proved theoretically, its convergence can be found experimentally in our simulation.

\bibliography{test}

\end{document}